\documentclass[conference]{IEEEtran}
\IEEEoverridecommandlockouts

\usepackage{amsmath,amssymb,amsfonts}
\usepackage{algorithmic}
\usepackage{graphicx}
\usepackage{epsfig}
\usepackage{epstopdf}
\usepackage{textcomp}
\usepackage{xcolor}
\usepackage{multirow}
\usepackage{amsmath}
\usepackage{subcaption}
\usepackage{float}
\usepackage{siunitx}
\usepackage{hyperref}
\usepackage{soul}
\usepackage{authblk}
\usepackage{colortbl}
\usepackage[style=ieee,dashed=false]{biblatex}
\hypersetup{colorlinks,urlcolor=blue}

\def\BibTeX{{\rm B\kern-.05em{\sc i\kern-.025em b}\kern-.08em
    T\kern-.1667em\lower.7ex\hbox{E}\kern-.125emX}}
\begin{document}

\title{OpenCV2X: Modelling of the V2X Cellular Sidelink and Performance Evaluation for Aperiodic Traffic}

\author[*]{Brian McCarthy}
\author[$\dag$]{Andres Burbano-Abril}
\author[$\dag$]{Victor Rangel Licea}
\author[*]{Aisling O'Driscoll}

\affil[*]{{\textit{School of Computer Science and Information Technology}, \textit{University College Cork}, Cork, Ireland}}
\affil[$\dag$]{\textit{School of Engineering}, \textit{National Autonomous University of Mexico}, Mexico City, Mexico}
\affil[*]{\{b.mccarthy, a.odriscoll\}@cs.ucc.ie}
\affil[$\dag$]{bburbano@comunidad.unam.mx, victor@fi-b.unam.mx}

\renewcommand\Authands{ and }

\maketitle
\begin{abstract}
This paper presents OpenCV2X, the first publicly available, open-source simulation model of the Third Generation Partnership Project (3GPP) Release 14 Cellular Vehicle to Everything (C-V2X) sidelink, which forms the basis for 5G NR Mode 2 under later releases. This model is fully compliant with the existing vehicular service and application layers, including messaging sets as defined by the automotive and standards communities providing a fully standardised, cross-layer communication model. Using this model, we show how the current sidelink scheduling mechanism performs poorly when scheduling applications with highly aperiodic communication characteristics, such as ETSI Cooperative Awareness Messages (CAMs). We then provide the first indepth evaluation of dedicated per-packet aperiodic scheduling mechanisms, in contrast to schemes that parameterise the existing algorithm. This paper highlights that the level of aperiodicity exhibited by the application model greatly impacts scheduling performance. Finally, we analyse how such scheduling mechanisms might co-exist. 
\end{abstract}


\section{Introduction}
\label{sec:intro}
\footnote{This work has been submitted to the IEEE for possible publication. Copyright may be transferred without notice, after which this version may no longer be accessible.}
In recent years the 3GPP have specified mobile standards to support V2X (vehicle to everything) communications and to compete with the existing wireless standards based on 802.11p. These standards, known as Release 14 \cite{3gpp-TR-36-885} and Release 15 \cite{3gpp-rel15} support vehicle to infrastructure/network (V2I/V2N) communications via the traditional \textit{Uu} interface but also allows for direct communication between vehicles (V2V) via the \textit{PC5} sidelink interface. The radio resources necessary to facilitate V2V communications can be selected and managed by the cellular network (Mode 3) or selected autonomously by the vehicles (Mode 4) using the distributed scheduling algorithm, Sensing Based Semi-Persistent Scheduling (SB-SPS). The specification in Rel. 14 and Rel. 15 has acted as a pre-cursor for the C-V2X New Radio (NR) specification in 3GPP Release 16 \cite{3gpp-rel16} with Mode 4 forming the basis for NR V2X Mode 2. There is no difference in their resource scheduling mechanism although Rel. 16 acknowledges the need for prioritisation mechanisms for aperiodic application traffic to be specified.  

The original focus of the 3GPP Rel. 14 standard was on traffic safety and efficiency applications with the assumption that CAMs would be shared periodically between vehicles.  This assumption does not typically hold true as ETSI messaging generation rules \cite{etsi-cam} specify CAM transmission based on vehicle dynamics. Furthermore, the 3GPP have also specified enhanced V2X (eV2X) applications for connected and automated vehicles including platooning, extended sensors, advanced and remote driving that will need to support aperiodic application traffic patterns. 

However, the transmission of aperiodic application traffic has a large impact on the operation of the SB-SPS MAC scheduling mechanism and was designed with periodicity in mind, thereby enabling accurate prediction of free radio resources. As such an open research challenge is to fully understand the cause of SB-SPS performance degradation when dealing with aperiodic application traffic and to redesign the C-V2X MAC radio resourcing algorithm to provide better support, either by modifying parameters within the existing SB-SPS algorithm or by implementing a separate prioritisation scheme, as identified by Rel 16 (or a combination thereof). 

As such, this study describes an open source system level simulator available to the vehicular communications community to deepen knowledge surrounding C-V2X performance. We further perform an in-depth study to determine the precise conditions under which SB-SPS exhibits significant performance degradation when supporting aperiodic application traffic across a variety of vehicular densities and application models. We observe that frequent grant breaks leads to inefficient use of radio resources and ultimately increased packet collisions. 

Uniquely, this paper present the first study in literature of the performance of dedicated the aperiodic prioritisation mechanisms for C-V2X and demonstrate that the level of aperiodicity i.e. packet inter-arrival rate, plays a significant role in the impact on SB-SPS performance. Specifically we consider applications models as specified by ETSI and 3GPP.  We find that application models with highly aperiodic characteristics need a dedicated prioritisation mechanism, whereas those that exhibit lower variation in packet inter-arrival rates can be accommodated by parameterisation within the existing SB-SPS algorithm.

The main contributions of this paper are:

\begin{itemize}
    \item A full stack, open source simulator of the C-V2X sidelink from the application layer right through the PHY layer that is fully compliant with ETSI messaging sets and European communication standards. The validated model is available at  \footnote{\url{http://www.cs.ucc.ie/cv2x/}}. 
    \label{page:footnote}
    \item An indepth study of the behaviour of the SB-SPS algorithm when faced with aperiodic application traffic characteristics of variable packet inter-arrival rates.
    \item An evaluation of the effectiveness of candidate scheduling mechanisms for aperiodic traffic, as suggested by 3GPP Rel. 16, in order to gain deeper insight into how aperiodic application traffic, characteristic of future e-V2X 5G NR applications, can be better supported. 
     \item A study of how dedicated aperiodic scheduling mechanisms co-exist with default SB-SPS for mixed application models. 
\end{itemize}

The remainder of the paper is organised as follows. Section \ref{lit review} describes the related literature with Section \ref{Mode4_desc} describing the operation of the PHY layer and the SB-SPS MAC layer algorithm of Rel. 14 (Mode 4), which is the basis for Rel. 16 (Mode 2). The devised model is described in Section \ref{sec_implementation}, with key implementation details outlined. Section \ref{sec:validation} validates the simulation model against an analytical model to verify correct operation of OpenCV2X. Section \ref{sec:aperiodic_traffic} provides a detailed simulation study of the challenges in supporting aperiodic application traffic, using the default SB-SPS resource scheduling mechanism for ETSI and 3GPP application models. Section \ref{sec:aperiodic_fixes} evaluates mechanisms to support aperiodic traffic, specifically those designed to operate within the existing SB-SPS algorithm, and those designed exclusively to support aperiodic traffic. The implications of these algorithms for mixed application models is also considered. Finally sections \ref{sec:discussion} and \ref{sec:conclusion} summarise the main findings and implications of this study and conclude the paper.

\section{Related Literature}
\label{lit review}

\subsection{Simulation Models}
\label{subsec:simmodels}

Existing models can be categorised as system level simulators or analytical models such as \cite{analytical,lte-v2vsim}, that are often MatLab based and only characterise the link level thereby limiting investigation. OpenCV2X has been fully validated against \cite{analytical} to ensure correctness of the model. Existing research published in \cite{Molina-Masegosa,roux-model} use custom system level simulators but these are not available to the wider vehicular communications community limiting reproducibility. To the best of the authors knowledge, there exists only 2 system level simulators \cite{ns3-model, Artery-c}, intended for contribution to the wider research community. \cite{ns3-model} is an open source, publicly available C-V2X mode 4 model built on the NS-3 simulator. In contrast to OpenCV2X, it does not support non-IP vehicular applications with the existing cellular stack, nor does it integrate with standardised automotive application and facilities layers to provide a fully compliant cross layer simulator. Artery-C \cite{Artery-c} is a recent addition, from the developers of Artery \cite{artery}. This is the closest simulation model to OpenCV2X in that it builds on the OMNeT++ simulator as well as the Artery and SimuLTE frameworks. Artery-C integrates with the ETSI application \& facilities layers, supports non IP based V2X applications and the autonomous sidelink. However at the time of writing this paper, Artery-C is not currently available as open source to the wider vehicular communications community, although the authors state that there are plans to do so.

\subsection{Characterisation of SB-SPS Performance for Aperiodic Traffic}
\label{subsec:aperiodic}

This paper uses OpenCV2X to perform an indepth study on the limitations of the default SB-SPS algorithm in supporting application traffic with aperiodic characteristics and assesses mechanisms to overcome these limitations. Past studies have stated that SB-SPS has been designed to better facilitate periodic traffic \cite{bazzi2020congestion,analytical}. However these studies have not provided indepth discussion on causation of declined performance nor quantified the diminished performance. Three papers have taken some steps towards studying this performance decline. 

Recently, Romeo et. al \cite{icc-sts} consider aperiodic traffic in the form of Decentralized Environmental Notification Messages (DENMs) which are sent aperiodically to alert vehicles of hazardous road conditions. The authors examine the impact of tuning SB-SPS parameters to support aperiodic packets e.g. by reducing sensing windows, selection windows and selection probability (\textit{RSel}) when providing CSRs to the MAC layer. One off single DENM packets are considered as opposed to a traffic pattern/model with aperiodic arrival rates. Their main finding is that when reducing the selection window to meet the lower latency requirements of DENMs, the PDR (Packet Delivery Rate) can be impacted. To address this, the authors reduce RSel to 10\% and 5\% which provides some improvement with the caveat of increasing collisions at close distances in denser scenarios. 
Additionally the authors explore different SB-SPS parameters to explicitly support DENM repetitions \cite{icc-repetition}, with repetition packets sent at periodic 100ms intervals.  Two approaches are considered; the first schedules the DENM + repetitions as per default SB-SPS (albeit with a reduced sensing window of 100ms); the second uses the keep probability parameter (\textit{probResourceKeep = 0}) to schedule a new resource for every packet, based on the sensing window i.e. creating a single one time 'grant'. This decreases the likelihood of two or more Vehicular UEs (VUEs) arbitrarily choosing the same resource, with the subsequent collision being maintained for the duration of the grant and as such increases the likelihood of successful receipt of the DENM. This work only considers periodic CAM transmissions and does not consider aperiodic ETSI CAMs in conjunction with DENMs. 

Lastly, Molina-Masegosa et. al \cite{aperiodic-molina} conducted a study contrasting C-V2X Mode 4 with 802.11p for periodic and aperiodic application models, as well as variable packet sizes. We believe that Molina et. al and this paper are the only studies of ETSI CAMs performance with C-V2X.  They further study adapting the resource reservation interval by balancing re-selections against occurrences of wasted resources using three strategies. However they assume that grant breaking is disabled and do not compare against a grant breaking mechanism as addressed in this paper. Notably the authors suggest that the SB-SPS mechanism is fundamentally counter-productive for aperiodic traffic and highlight that further dedicated schemes are required.

\subsection{Prioritisation Mechanisms for Aperiodic Traffic}
\label{subsec:priorMechs}

Notably of the identified studies, none investigate prioritisation mechanisms for aperiodic traffic, which has been identified as necessary for 5G NR V2X in Release 16 and 17. Such an investigation is vital to understanding which features will best support future vehicular applications, and as such should be adopted by future standards and addressed in the literature. Such a study is currently absent from literature, but this paper takes steps to address this. When identifying candidate mechanisms, Rel. 16 refers to a number of industry led proposals, including Listen Before Talk by Qualcomm \cite{Qualcomm-meeting}, a counter-based mechanism by LG \cite{lg-counter} and short-term reservations (STR) by Ericsson \cite{ericsson-one-shot}. These proposals, in turn cite a report by Intel examining reduced sensing windows \& removal of RSSI filtering \cite{intel-no-rssi-initial}. This paper is the first to evaluate their performance in literature. 

\section{Operation of the C-V2X Sidelink}
\label{Mode4_desc}

C-V2X Mode 4 introduces modifications to the PHY and MAC layers of the LTE sidelink to support V2X communications. The PHY layer is designed to improve the performance of LTE under high mobility conditions. At the MAC layer, a new scheduling mechanism (SB-SPS) is implemented to allow vehicles to select resources autonomously. The following sections describe the most important aspects of the PHY and MAC layers of C-V2X Mode 4.

\subsection{C-V2X Physical Layer}
\label{subsec:phylayer}

The PHY layer of C-V2X implements Single-Carrier Frequency Division Multiple Access (SC-FDMA). In the time domain, resources are organized into subframes of 1 ms, which are further grouped into frames of 10 ms. Each subframe contains 14 SC-FDMA symbols, 4 symbols are used for demodulation reference signals (DMRS), 1 symbol for Tx-Rx switching, and the remaining 9 symbols are left for data transmissions.

In the frequency domain, the channel is divided into subcarriers of 15 kHz. These subcarriers are grouped into Resource Blocks (RBs), with each RB containing 12 subcarriers and spanning over 1 subframe. Unlike the conventional resource structure of LTE, C-V2X groups RBs into subchannels. The number of RBs per subchannel and the number of subchannels are configurable but limited by the allocated bandwidth, which can be of 10 or 20 MHz.

Two physical channels exist; the Physical Sidelink Shared Channel (PSSCH) and Physical Sidelink Control Channel (PSCCH). The PSSCH transmits the RBs carrying data, also known as Transport Blocks (TBs). The PSCCH carries the Sidelink Control Information (SCI), which is critical for scheduling and decoding. The SCI contains information such as the Modulation and Coding Scheme (MCS) used to transmit the packet, the frequency resource location of the transmission, and other scheduling information.

The PSCCH and PSSCH can be transmitted using adjacent or non-adjacent schemes. In the adjacent scheme, the PSCCH and PSSCH are transmitted in contiguous RBs. Differently, in the non-adjacent scheme, the PSCCH and PSSCH are transmitted in different RBs pools. In terms of occupancy,  the PSCCH requires 2 RBs, while the number of RBs required by the PSSCH is variable and depends on the size of the TB. It is worth noting that the PSSCH and PSCCH are always transmitted in the same subframe independently of the transmission scheme.

\subsection{C-V2X Medium Access Control Layer}
\label{subsec:maclayer}
At the MAC layer, C-V2X implements SB-SPS, dubbed Mode 4, to allow vehicles to select resources autonomously. The process starts with the reception of a packet from the upper layers. Upon reception, the MAC layer creates a scheduling grant containing the number of subchannels, the number of recurrent transmissions for which the subchannels will be reserved, and the periodicity between transmissions. If a grant has already been created at the time a packet is received from the upper layers, the transmission is scheduled using the existing grant. The number of subchannels is pre-configured and depends on the application requirements. The number of recurrent transmissions is defined by the Resource Reselection Counter (RRC), which is set by randomly selecting an integer value between 5 and 15. Finally, the periodicity between transmissions is defined by the Resource Reservation Interval (RRI), whose value is set by upper layers. 

The grant is then passed to the PHY layer, which generates a list of all the subchannels meeting the grant specifications. These subchannels are known as Candidate Single-Subframe Resources (CSRs) and consist of one or multiple subchannels in the same subframe. The list contains all the CSRs within a selection window comprising the period between the time the packet is received from the upper layers and the maximum allowed latency defined by the RRI. 

The list is then filtered using the information in the SCIs received during a sensing window comprised of the last 1000 subframes. Based on this information, CSRs are excluded if the SCI indicates the CSR will be reserved during the upcoming selection window and if the average PSSCH Reference Signal Received Power (RSRP) of the CSR exceeds a predefined threshold. After excluding all CSRs that meet these two conditions, at least 20\% of all the CSRs should remain available. If this is not the case, the process is repeated by increasing the RSRP threshold by 3dB. Finally, the PHY selects the 20\% of CSRs with the lowest Sidelink Reference Signal Strength Indicator (RSSI) averaged over the sensing window. This ensures the CSRs with the lowest levels of interference are considered for selection. 

The remaining CSRs are passed to the MAC layer, where a single CSR is selected at random to reduce the probability of multiple vehicles choosing the same CSR. The CSR is selected for a number of recurrent transmissions defined by the RRC, whose value is decreased by one after each transmission. When the RRC value reaches zero, each vehicle can maintain the same reservation with probability \textit{P} or generate a new grant and restart the selection process. The value of \textit{P} is configured previously and can take any value between [0,0.8].

\section{Model implementation}
\label{sec_implementation}
The implemented model is open-source and enables the vehicular communications research community to evaluate the performance of their hypotheses, using an end-to-end full-stack approach, thereby stimulating further research on this topic. It further allows for full reproducibility of the discussed results. In the following sections, we describe the model architecture as well as the main implementation aspects of the PHY and MAC layers of C-V2X Mode 4. 

\subsection{OpenCV2X Model architecture}
\label{subsec:modelarchitecture}

\begin{figure}[htbp]
\centering
\includegraphics[scale=0.4]{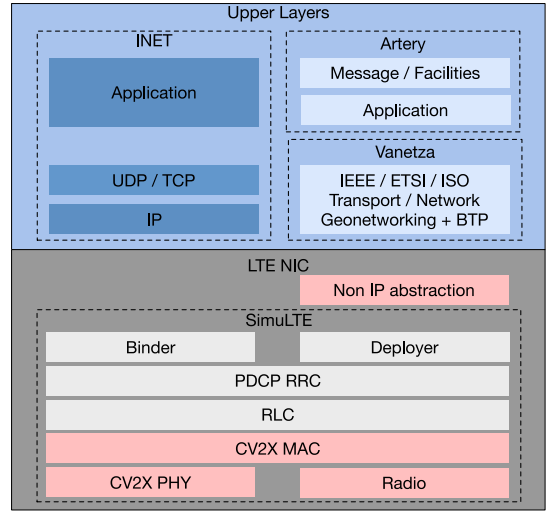}
\caption{OpenCV2X model architecture.}
\label{fig_architecture}
\end{figure}

Fig. \ref{fig_architecture} highlights the OpenCV2X stack. It can be seen that OpenCV2X models the full cellular stack from application layer right through the radio for the VUE in Mode 4. In the proposed architecture, upper layers can leverage either the \textit{INET} or \textit{Artery} \cite{artery} frameworks. The protocol stack of \textit{INET} can be readily integrated in \textit{Omnet++} with a generic application on top of it. Of most relevance to this work, \textit{Artery} provides the implementation of the ETSI ITS-G5 protocol stack. This enables simulation of applications such as CAMs and DENMs, as well the GeoNetworking and Basic Transport Protocol (BTP) protocols. Importantly, OpenCV2X was designed such that it is also possible to use the model in a standalone state, that does not require \textit{Artery} or the facilities/application layers if not required, and instead operates with the \textit{Veins} network model for integration with the SUMO road network simulator. This offers users of OpenCV2X increased flexibility in which vehicular application they wish to model on top of the C-V2X radio.  

For the lower layers, our model extends the \textit{SimuLTE} framework \cite{simulte}, which provides a system-level model for LTE. Recognising that many vehicular applications will not be IP based, a non-IP interface was added to allow Artery to communicate with the lower C-V2X layers and \textit{SimuLTE}. To model the specifics of the C-V2X Mode 4 operation, we introduced new MAC and PHY layers. The integration of \textit{Artery} and \textit{SimuLTE} also required the implementation of a new radio driver and Network Interface Card (NIC). Moreover, important concepts in \textit{SimuLTE}, such as the Binder and Deployer were modified to operate as standalone modules to allow UEs in the simulation without the need of an eNB. The major new features are shown in Fig. \ref{fig_architecture} in pink, although some minor changes were required at other layers for integration. Interested readers can refer to \cite{openCV2XFirst} for further details or visit the OpenCV2X website \pageref{page:footnote}.

\subsection{OpenCV2X PHY and MAC layers implementation}
\label{subsec:phymacImplementation}

To model the PHY and MAC layers of C-V2X, two new classes, \textit{LteMacVueMode4} and \textit{LtePhyVueMode4}, were introduced. These new classes replace the \textit{LteMacUeD2D} and \textit{LtePhyUeD2D} classes in \textit{SimuLTE} based on 3GPP Release 12 D2D Proximity-based Services (ProSe) \cite{3gpp-rel12}. These new classes were required to model the specific operation of the C-V2X mode 4 sidelink, including autonomous resource allocation. 

The MAC layer is implemented in the \textit{LteMacVueMode4}. Here, a packet from the RLC layer either triggers the generation of a new grant or the scheduling of a packet transmission if a grant has already been allocated. In turn, \textit{LtePhyVueMode4} implements the operation of the SB-SPS mechanism, including the sensing and filtering processes required to select CSRs. 

\textit{LtePhyVueMode4} also leverages the \textit{LteRealisticChannelModel} class in \textit{SimuLTE} to model the transmission channel in C-V2X. To comply with the simulation guidelines specified by 3GPP in \cite{3gpp-TR-36-885}, we modified \textit{LteRealisticChannelModel} by implementing the WINNER+ B1 channel model and the Block Error Rate curves described in \cite{nist-blers}. Moreover, we modified the computation of power levels and implemented new physical layer parameters, specifically required by C-V2X mode 4 and defined in the 3GPP release 14 standard. The following section describes these modifications in detail.

\subsection{OpenCV2X PHY layer extensions}
\label{subsec:validationChanges}

A number of significant changes are necessary in the SimuLTE PHY layer to support C-V2X sidelink mode 4. Specifically:

\begin{enumerate}
    \item PHY layer parameterisation as a consequence of having power levels distributed across transmitted RBs.
    \item Error rate calculation on a per packet basis.
\end{enumerate}

In the current version of the SimuLTE \textit{LteRealisticChannelModel}, packets are transmitted as groups of RBs and every RB is allocated the total transmission power. This transmission model is generally accurate as long as the transmission bandwidth is maintained fixed during the simulation and all power ratios are clearly defined. Furthermore, in SimuLTE, the receiver transmissions are evaluated on an RB basis by computing the SINR of each RB and mapping the result to a BLER value. Importantly, all interferers are assumed to use the total transmission power for each RB and the thermal noise is set as a function of the total bandwidth.

However, in C-V2X Mode 4 the transmission bandwidth can be highly variable due to the variety of subchannel configurations, MCS' and packet sizes. As a result, OpenCV2X implements a solution where rather than every RB being allocated the total transmission power, it distributes all power levels i.e. the transmission power, interference and thermal noise, across the number of RBs required to transmit each packet. This is vital to the correct calculation of key C-V2X-specific physical layer parameters, namely PSSCH-RSRP and S-RSSI since both parameters depend directly on the number of transmitted RBs \cite{3gppMeasurements}. Additionally, it impacts the correct calculation of PHY layer measurements such as SINR. It is worth noting that PSSCH-RSRP and S-RSSI are not implemented in \textit{SimuLTE}, since the current version provides a model for network-assisted D2D communication as specified by the 3GPP release 12 standard \cite{simulte-D2D}.

To determine the number of RBs for each packet transmission, we consider both the selected MCS and the size of the packet to obtain the TB size from Table 7.1.7.2.1 in \cite{3gppPhyProcedures}. The TB size defines the number of RBs required to transmit a packet considering the Cyclic Redundancy Check (CRC), payload and redundancy bits. In addition, since we consider the case of adjacent PSCCH+PSSCH transmissions, the 2 RBs needed to transmit the SCI are added to obtain the total number of RBs to be transmitted. At the receiver, the total received power (after accounting for pathloss and shadowing) is divided by the number of transmitted RBs to obtain the received power per RB. This is then divided by the transmission bandwidth to obtain the Power Spectral Density (PSD) of each RB. The PSD is the building block on which key C-V2X PHY layer measurements such as S-RSSI, PSSCH-RSRP and SINR are computed.

To compute the PSSCH-RSRP, the PSD of the RB is multiplied by the bandwidth of each Resource Element (RE). We assume the PSD to be constant within each RB and, therefore, the power per RE to be the same for all REs. With this assumption, the PSSCH-RSRP, defined as the average power contribution of each RE, is equal to the power per RE. For the S-RSSI, the received power per SC-FDMA symbol is obtained. To do this, we consider all power sources, i.e. the power, interference, and noise per RB. The received power per SC-FDMA symbol is then computed as the sum of the total received power per RB multiplied by the total number of transmitted RBs. 

Finally, we compute the SINR of each RB as the ratio between the received power and the power of the interference plus the total noise, all on an RB scale. To compute interference we account for the PSD per RB from all transmitters using overlapping RBs during an ongoing transmission. By using the PSD per RB and the number of RBs used by each interferer, we can accurately compute the interference. The total noise is also estimated based on the number of transmitted RBs, by adding the noise figure at the receiver to the thermal noise. The thermal noise is computed by multiplying the thermal noise density specified in the LTE standard of -174 dBm/Hz \cite{3gpp-noise} by the number of transmitted RBs and the RB bandwidth. 

The second significant change that OpenCV2X incorporates relates to the calculation of error rates. OpenCV2X calculates error rates on a per packet basis in contrast to SimuLTE, where transmissions are evaluated per RB by computing the SINR of each RB and mapping the result to a BLER value. As SimuLTE computes the SINR by assuming that all interferers use the total transmission power for each RB, and the thermal noise is set as a function of the total bandwidth. This would result in incorrect C-V2X PHY calculations. In contrast, OpenCV2X inputs the mean SINR for the transmitted RBs into the Mode 4 BLER curves defined in \cite{nist-blers} to obtain the packet error rate using the curve that corresponds to the MCS of each transmission.

\section{OpenCV2X Validation}
\label{sec:validation}

\begin{figure}[htbp]
\begin{subfigure}{.48\textwidth}
  \centering
  \includegraphics[width=.82\linewidth]{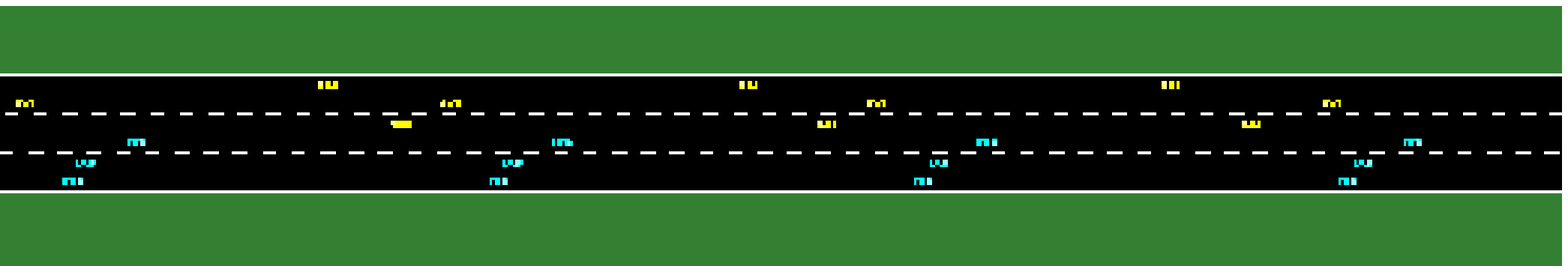}
  \caption{Density of $\beta$=0.06 veh/m.}
  \label{fig:road-0.06}
\end{subfigure}
\begin{subfigure}{.48\textwidth}
  \centering
  \includegraphics[width=.82\linewidth]{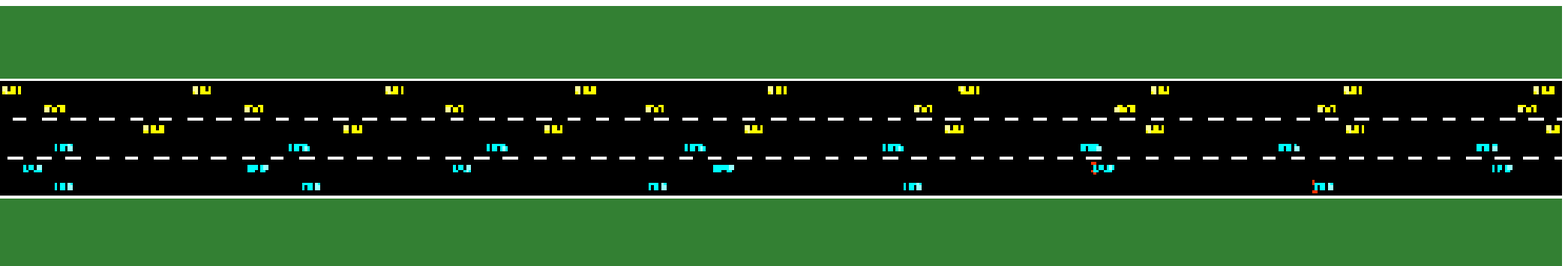}
  \caption{Density of $\beta$=0.12 veh/m.}
  \label{fig:road-0.12}
\end{subfigure}
\begin{subfigure}{.48\textwidth}
  \centering
  \includegraphics[width=.82\linewidth]{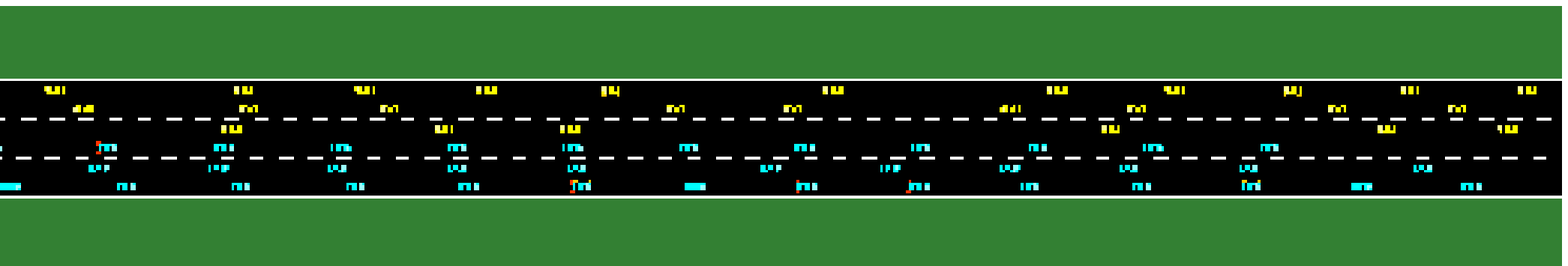}
  \caption{Density of $\beta$=0.2 veh/m.}
  \label{fig:road-0.2}
\end{subfigure}
\begin{subfigure}{.48\textwidth}
  \centering
  \includegraphics[width=.82\linewidth]{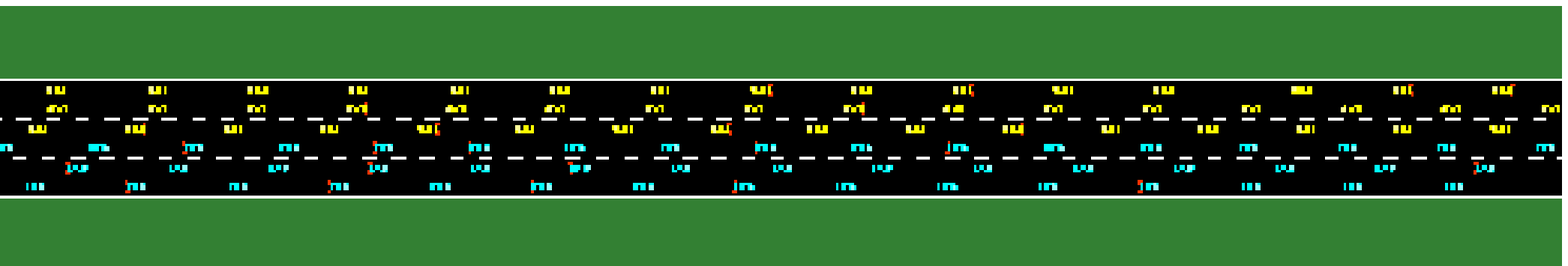}
  \caption{Density of $\beta$=0.3 veh/m.}
  \label{fig:road-0.3}
\end{subfigure}
\caption{Vehicular densities.}
\label{fig:roads}
\end{figure}

To validate the OpenCV2X model and perform a detailed study of the performance of aperiodic application traffic, OpenCV2X is used in conjunction with SUMO, which acts as a road network simulator and provides VUE mobility. The road environment and vehicular densities considered are summarised in Table \ref{tab:road}.,and can also be seen in Fig. \ref{fig:roads} to provide a visual comparison of the various densities. Other key simulation parameters are summarised in Table \ref{tab:cv2x-setup}. 

\begin{table}[htbp]
\caption{Vehicular Environment.}
\label{vehicular scenarios}
\begin{center}
\begin{tabular}{l l}
\hline\hline 
\textbf{Parameter} & \textbf{Value} \\ [0.5ex] 
\hline
Vehicular density & 0.06, 0.09, 0.12, 0.16,  0.2, 0.25 and 0.3 veh/m\\

Road length & 2 km \\

Number of lanes & 3 in each direction (6 in total)\\

Lane width & 4 m \\

SUMO step-length & 1ms\\
\hline\hline
\end{tabular}
\label{tab:road}
\end{center}
\end{table} 

\begin{table}[htbp]
\caption{Simulation parameters.}
\begin{center}
\begin{tabular}{l l}
\hline\hline
\textbf{Parameter} & \textbf{Value}\\
\hline\hline
\multicolumn{2}{c}{\textbf{Channel settings}}\\
\hline
Carrier frequency & 5.9 GHz\\
Channel bandwidth & 10 MHz\\
Number of subchannels & 2 \\
Subchannel size & 12 Resource Blocks\\
Adjacency mode PSCCH-PSCCH & Adjacent\\
\hline
\multicolumn{2}{c}{\textbf{Application layer}}\\
\hline
Packet size & 190 Bytes\\
Transmission frequency ($F_{Tx}$) & 10 Hz\\
Artery Middleware Update Interval & 1ms \\
\hline
\multicolumn{2}{c}{\textbf{MAC \& PHY layer}}\\
\hline
Resource keep probability & 0\\
RSRP threshold & -126 dBm \\
RSSI threshold & -90 dBm \\
Propagation model & Winner+ B1\\
MCS & 7 (QPSK 0.5)\\
Transmission power ($P_{Tx}$) & 10, 20 and 23 dBm\\
Noise figure & 9 dB\\
Shadowing variance LOS & 3 dB\\
\hline\hline
\end{tabular}
\label{tab:cv2x-setup}
\end{center}
\end{table}

To validate our model implementation, we benchmark against an analytical LTE-V Mode 4 model made available by Gonzalez-Martin et. al \cite{analytical} in Matlab. We examine Mode 4 performance while observing parameters that impact the communication performance such as vehicular density, transmission power levels and the MCS. 

\begin{figure}[htbp]
\centerline{\includegraphics[width=.82\linewidth]{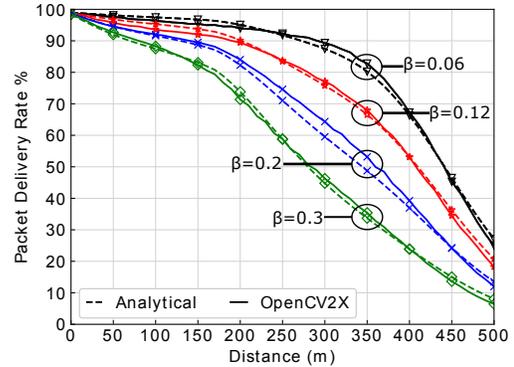}}
\caption{PDR as a function of the distance between the transmitter and receiver for varying vehicular densities.}
\label{fig_density}
\end{figure}

Fig. \ref{fig_density} shows the Packet Delivery Rate (PDR) performance of the OpenCV2X simulator (solid lines) when compared to the analytical model (dashed lines). This considers \textit{P\textsubscript{Tx}}=20dBm, \textit{F\textsubscript{Tx}}=10Hz, MCS 7 (QPSK 0.5, 2 subchannels) and four vehicular densities, \(\beta\)=0.06, 0.12, 0.2 and 0.3 veh/m. The lower vehicular densities were chosen to comply with recommended 3GPP C-V2X simulation guidelines, specifically the Highway Fast and Slow scenarios \cite{3gpp}, while the higher density scenarios closely compare with the performance of the analytical model \cite{analytical}. The 3GPP Highway scenarios result in a mean vehicular density of \(\beta\)=0.06 and 0.12 veh/m respectively. We disregard the first 500 seconds of the simulation to discount vehicles entering the simulation. 

It can be observed in Fig. \ref{fig_density} that the simulated PDR closely aligns with the  analytical model with only a small divergence in PDR, irrespective of the vehicular density considered. Table \ref{table_mad_pdr} considers further densities to validate the accuracy of the simulation model, by examining the mean absolute deviation in PDR when compared to the analytical estimate. In all cases, it can be seen that it differs by less than 2\%. 

\begin{table}[htbp]
\caption{Mean absolute deviation in PDR between OpenCV2X and the analytical model for varying vehicular densities.}
\begin{center}
\begin{tabular}{|c|c|c|c|}
\hline
$P_{TX}$ (dBm) & $F_{TX}$ (Hz) & $\beta$ (veh/m) & PDR (\%)\\
\hline
\multirow{7}{*}{20} & \multirow{7}{*}{10} & 0.06 & 1.20\\\cline{3-4}
\multirow{7}{*}{} & \multirow{7}{*}{} & 0.09 & 1.35\\\cline{3-4}
\multirow{7}{*}{} & \multirow{7}{*}{} & 0.12 & 1.15\\\cline{3-4}
\multirow{7}{*}{} & \multirow{7}{*}{} & 0.16 & 1.21\\\cline{3-4}
\multirow{7}{*}{} & \multirow{7}{*}{} & 0.2 & 1.84\\\cline{3-4}
\multirow{7}{*}{} & \multirow{7}{*}{} & 0.25 & 1.48\\\cline{3-4}
\multirow{7}{*}{} & \multirow{7}{*}{} & 0.3 & 1.07\\\hline
\end{tabular}
\end{center}
\label{table_mad_pdr}
\end{table} 

\begin{figure}[htbp]
\begin{subfigure}{.48\textwidth}
  \centering
  \includegraphics[width=.82\linewidth]{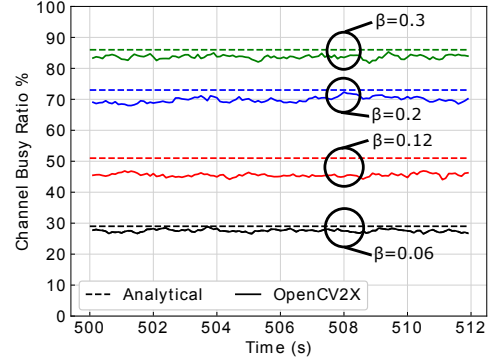}
  \caption{PSSCH Channel Busy Ratio (CBR) for varying vehicular densities.}
  \label{fig:sfig_cbr_line}
\end{subfigure}
\begin{subfigure}{.48\textwidth}
  \centering
  \includegraphics[width=.82\linewidth]{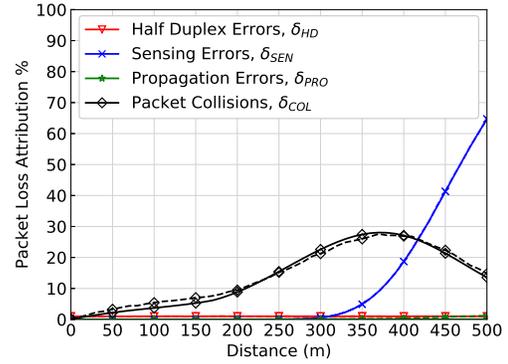}
  \caption{Packet Loss Attribution per type of error ($\beta$=0.12 veh/m) for OpenCV2X (solid lines) and the analytical model (dashed lines).}
  \label{fig:sfig_error}
\end{subfigure}
\caption{Channel conditions including (a) CBR \& (b) packet loss attribution.}
\label{fig:baseline}
\end{figure}

To ensure that the simulated and analytical models are considering comparable channel characteristics and behaviour, we further examine the channel load by measuring the Channel Busy Ratio (CBR), which is the number of subchannels sensed as occupied, as well as the causes of packet loss. In Fig. \ref{fig:sfig_cbr_line}, we can see that both models consider comparable channel loads with a mean deviation up to 6\%. It should be noted that the analytical model produces a single CBR estimate. Occupied subchannels are identified as those filtered for exceeding the \textit{RSRP} threshold with a mean CBR value subsequently estimated for all vehicles in the simulation. In contrast, OpenCV2X calculates CBR according to the 3GPP definition \cite{3gppMeasurements}, with each vehicle determining the percentage of subchannels in the last 100ms that have recorded an \textit{RSSI} greater than a pre-configured threshold of -90dBm. This, along with minor differences in the distribution of vehicular density, accounts for the minor divergence in channel load. In Fig. \ref{fig:sfig_error}, we investigate the causes of packet loss for the $\beta$=0.12 scenario. As outlined in \cite{analytical}, packet loss can be attributed to the following, and for clarity of comparison we use the same notation:
\ \\
\begin{itemize}
    \item Half Duplex Errors, \textit{\(\delta\)\textsubscript{HD}}: Vehicles receive a packet on the same subframe on which they are transmitting and are thus unable to decode the incoming packet.
    \item Sensing Errors, \textit{\(\delta\)\textsubscript{SEN}}: A received signal does not have sufficient power to meet the sensing power threshold  due to attenuation caused by path loss and shadowing.
    \item Propagation Errors, \textit{\(\delta\)\textsubscript{PRO}}: A propagation error occurs when the received power of the transmitted packet is higher than the sensing power threshold but the Signal to Noise Ratio (SNR) is insufficient to correctly decode the packet. 
    \item Packet Collisions, \textit{\(\delta\)\textsubscript{COL}}: Occurs when another vehicle transmits on the same subframe and subchannel; this interference impacts reception at the receiver by reducing the SINR.
\end{itemize}
\  \\
Notably, for all cases in Fig. \ref{fig:sfig_error}, the models aligned with little divergence. Similarly in Table \ref{tab:packetLoss}, the mean absolute deviation in packet loss errors between the models is considered, for varying vehicular densities. \textit{\(\delta\)\textsubscript{HD}} errors account for only 1\% of all packet loss in both models and incur 0.06\% divergence in such cases. \textit{\(\delta\)\textsubscript{SEN}} errors are based on the power of transmission and the attenuation incurred by the packet due to pathloss. It is a probability based mechanism which was implemented to match the analytical model and as such the divergence is less than 0.1\%. Based on the calculated attenuation, the received packet is only decoded once it exceeds the sensing power threshold of -90.5dbm. As expected these become the dominant cause of packet loss at greater distances. \textit{\(\delta\)\textsubscript{PRO}} errors are negligible in the simulation and of those that do occur there is minimal divergence from the analytical model of less than 0.36\% in all cases.

\textit{\(\delta\)\textsubscript{COL}} errors are the other dominant source of packet loss. These occur more frequently as the distance between the transmitter and receiver increases, as a result of increased attenuation. At distances exceeding 400m, these are eclipsed by sensing errors as the incoming packets do not reach the sensing power threshold. A marginal divergence of less than 2\% was observed for all evaluated scenarios as shown in Table \ref{tab:packetLoss}. Divergence is attributed to slightly lower CBR in OpenCV2X as observed in Fig. \ref{fig:sfig_cbr_line}, resulting in slightly less collisions and a minor improvement in PDR. This is attributable to the vehicular density and mobility. While the mobility scenarios were parameterised to have a density corresponding to the analytical model, the OpenCV2X model does not have a fixed density and as such it is not completely uniform. Furthermore we devised these experiments to comply with the 3GPP Highway road network comprising 6 lanes (3 in each direction) whereas Gonzalez-Martin et. al considered a 4 lane highway. As such, the density and subsequently the CBR considered in the analytical model are marginally higher. 

\begin{table}[htbp]
\caption{Mean absolute deviation in packet loss attribution between OpenCV2X and the analytical model for varying vehicular densities.}
\begin{center}
\resizebox{\columnwidth}{!}{%
\begin{tabular}{|c|c|c|c|c|c|c|}
\hline
$P_{TX}$(dBm) & $F_{TX}$(Hz) & \(\beta\) (veh/m) & \textit{\(\delta\)\textsubscript{HD}} (\%) & \textit{\(\delta\)\textsubscript{SEN}} (\%) & \textit{\(\delta\)\textsubscript{PRO}} (\%) & \textit{\(\delta\)\textsubscript{COL}} (\%)\\
\hline
\multirow{6}{*}{20} & \multirow{6}{*}{10} & 0.06 & 0.05 & 0.08 & 0.36 & 1.12 \\
\multirow{6}{*}{} & \multirow{6}{*}{}     & 0.09 & 0.05 & 0.05 & 0.22 & 1.20 \\
\multirow{6}{*}{} & \multirow{6}{*}{}     & 0.12 & 0.03 & 0.04 & 0.19 & 0.98 \\
\multirow{6}{*}{} & \multirow{6}{*}{}     & 0.16 & 0.02 & 0.04 & 0.16 & 1.12 \\
\multirow{6}{*}{} & \multirow{6}{*}{}     & 0.2  & 0.02 & 0.02 & 0.13 & 1.80 \\
\multirow{6}{*}{} & \multirow{6}{*}{}     & 0.25 & 0.02 & 0.02 & 0.09 & 1.44 \\
\multirow{6}{*}{} & \multirow{6}{*}{}     & 0.3  & 0.02 & 0.02 & 0.07 & 1.01 \\\hline
\end{tabular}%
}
\label{tab:packetLoss}
\end{center}
\end{table}

\begin{figure}[htbp]
\begin{subfigure}{.48\textwidth}
  \centering
  \includegraphics[width=.82\linewidth]{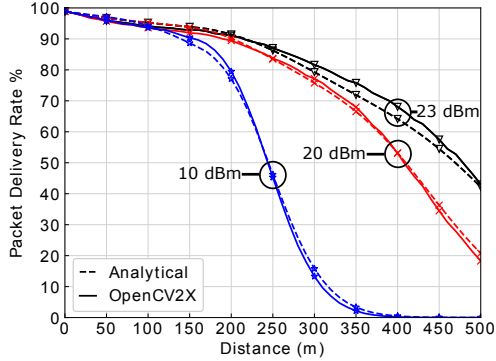}
  \caption{PDR as a function of distance for variable transmission power.}
  \label{fig:sfig_power}
\end{subfigure}
\begin{subfigure}{.48\textwidth}
  \centering
  \includegraphics[width=.82\linewidth]{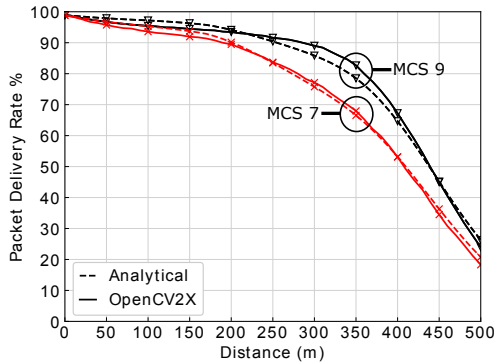}
  \caption{PDR as a function of distance for variable MCS.}
  \label{fig:sfig_mcs}
\end{subfigure}
\caption{Impact of varying transmission power and the modulation \& coding scheme on PDR.}
\end{figure}

We next considered the impact of transmission power and MCS on the PDR as shown in Fig. \ref{fig:sfig_power}. 
Three transmission powers \textit{P\textsubscript{Tx}}=10dBm, 20dBm and 23dBm, have been considered. 10dBm and 23dBm are the minimum and maximum transmission power limits specified by the 3GPP for C-V2X sidelink communications with 23dBm used as the default. 20dBm is considered for comparative purposes with \cite{analytical}. The ETSI ITS specification \cite{etsi-cam} includes a minimum transmission power of 10dBm to alleviate interference with CEN DSRC tolling systems. For all \textit{P\textsubscript{Tx}} values, the results are closely aligned with only negligible differences. As expected for the lowest value of \textit{P\textsubscript{Tx}}, the PDR is impacted significantly by reduced radio range as the distance between the transmitter and receiver increases. 

In Fig. \ref{fig:sfig_mcs} the MCS impact is considered, with MCS 7 (2 subchannels of size 14 RBs) and MCS 9 (4 subchannels of 12 RBs) evaluated. This assumes a vehicular density of $\beta$=0.12. Minimal divergence between the two models is observed. Using a higher MCS reduces the resources required to send a packet and can be a means of reducing channel congestion thereby improving PDR. The trade off is that such transmissions are less robust and more prone to loss. As such, it would be interesting to investigate the potential advantage of using adaptive MCS' and subchannel configurations as a means of C-V2X congestion control. 
\vspace{-1\baselineskip}

\section{SB-SPS Performance for Aperiodic Traffic}
\label{sec:aperiodic_traffic}

In this section we investigate how SB-SPS Mode 4 performs when application traffic follows an aperiodic arrival rate. For the remainder of the   experiments we consider \textit{P\textsubscript{Tx}}=23dBm, MCS 6, 3 subchannels of 16RBs each in accordance with 3GPP \cite{3gpp-TR-36-885} and four vehicular densities, \(\beta\)=0.06, 0.12, 0.2 and 0.3 veh/m with packet sizes  of 190 Bytes.
Packet sizes and latency requirements remain constant throughout all experiments to allow for direct comparison of the application models.  Application traffic is modelled as follows:
\begin{itemize}
    \item \textbf{Periodic (Default):} Static CAM transmission every 100ms with a default sensing window of 1000ms.
    \item \textbf{Aperiodic (3GPP):} Application layer packets arrive every 50ms + an exponential distribution with a mean of 50ms  \cite{3gpp-TR-37-885}.
    \item \textbf{Aperiodic (ETSI):} Importantly, previous literature often assumes a periodic CAM transmission rate. This does not align with the aperiodic transmission of CAMs according to the ETSI specification \cite{etsi} where CAMs are triggered based on a vehicle's dynamics i.e. deviation in heading (\textgreater4\textdegree), position (\textgreater4m) and speed (\textgreater0.5m/s) or at 1s intervals if these conditions are not satisfied.
    \item \textbf{Periodic Single Slot Usage:} Application traffic is transmitted every 100ms but the SB-SPS grant is deliberately broken after a single reservation as this represents the most inefficient use of SB-SPS allocated resources and can be considered as a worst case scenario.
\end{itemize}

\begin{figure}[htbp]
\begin{subfigure}{.48\textwidth}
  \centering
  \includegraphics[width=.82\linewidth]{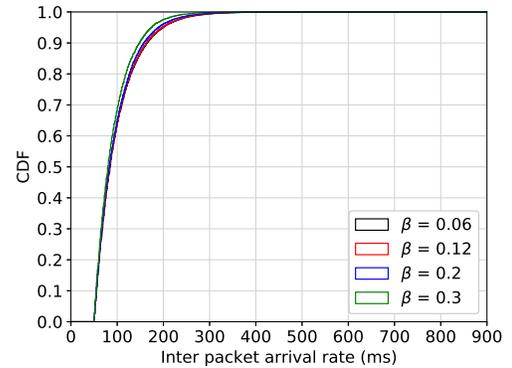}
  \caption{Packet inter-arrival rates for 3GPP.}
  \label{fig:interPacketArrival3GPP}
\end{subfigure}
\begin{subfigure}{.48\textwidth}
  \centering
  \includegraphics[width=.82\linewidth]{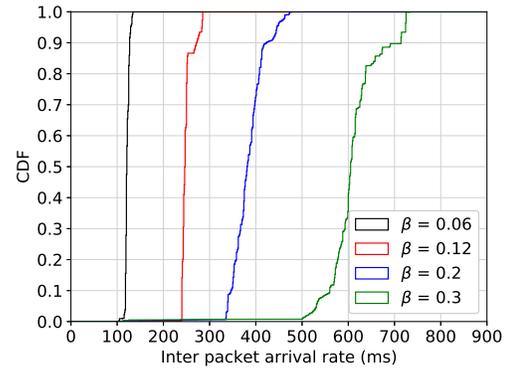}
  \caption{Packet inter-arrival rates for ETSI.}
  \label{fig:interPacketArrivalETSI}
\end{subfigure}
\caption{CDF of packet inter-arrival rates for aperiodic application models.}
\label{fig:interPacketArrival}
\end{figure}

To characterise the application models, CDFs of the packet inter-arrival rates are shown in Fig. \ref{fig:interPacketArrival}. In the 3GPP model, close to 100\% of the packet inter-arrival times are below 200ms, equivalent to 2 RRIs. The ETSI packet inter-arrival times are dependent on vehicle dynamics and thus increase with the vehicular density. This increase in packet inter-arrival times is important and is discussed in further detail later in this section.

\vspace{-1\baselineskip}

\begin{figure}[htbp]
\vspace{-.5\baselineskip}
\begin{subfigure}{.48\textwidth}
  \centering
  \includegraphics[width=.82\linewidth]{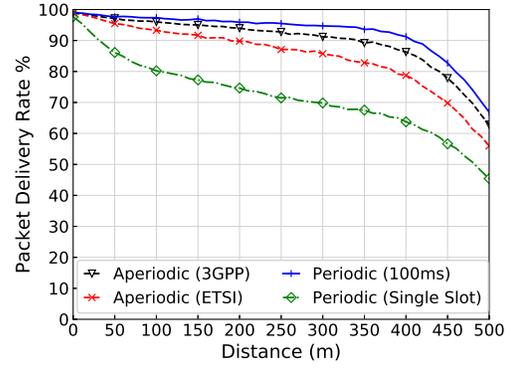}
  \vspace{-.25\baselineskip}
  \caption{\(\beta\)=0.06 veh/m.}
  \label{fig:low-scheduling-pdr}
\end{subfigure}
\vspace{-.15\baselineskip}
\begin{subfigure}{.48\textwidth}
  \centering
  \includegraphics[width=.82\linewidth]{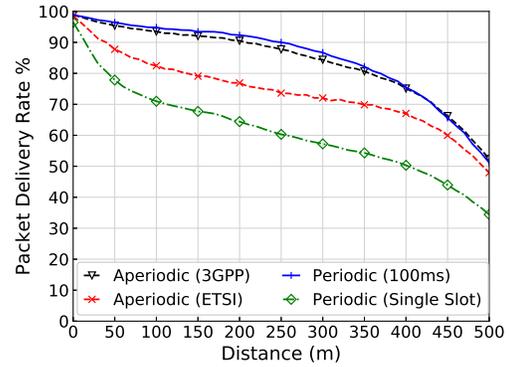}
  \vspace{-.15\baselineskip}
  \caption{ \(\beta\)=0.12 veh/m.}
  \label{fig:medium-scheduling-pdr}
\end{subfigure}
\vspace{-.15\baselineskip}
\begin{subfigure}{.48\textwidth}
  \centering
  \includegraphics[width=.82\linewidth]{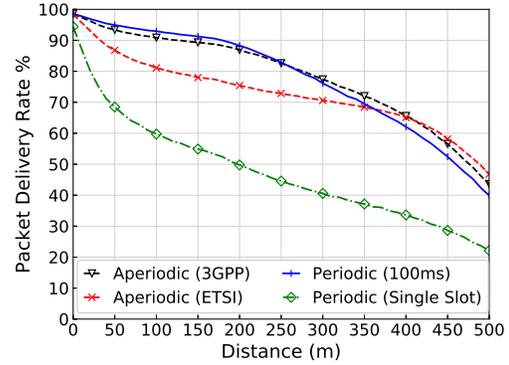}
  \vspace{-.25\baselineskip}
  \caption{\(\beta\)=0.2 veh/m.}
  \label{fig:high-scheduling-pdr}
\end{subfigure}
\vspace{-.15\baselineskip}
\begin{subfigure}{.48\textwidth}
  \centering
  \includegraphics[width=.82\linewidth]{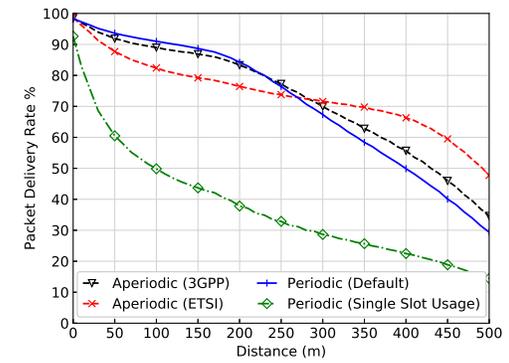}
  \vspace{-.25\baselineskip}
  \caption{\(\beta\)=0.3 veh/m.}
  \label{fig:highest-scheduling-pdr}
\end{subfigure}
\caption{PDR as a function of distance for periodic and aperiodic application models.}
\label{fig:scheduling-pdr}
\end{figure}

Fig. \ref{fig:scheduling-pdr} compares the PDR performance of these application models. In all cases it can be seen that the SB-SPS performance is compromised when supporting aperiodic traffic or traffic where the grants are frequently broken. Aperiodic 3GPP application traffic decreases PDR up to 5\% when compared with periodic traffic, though as the density increases the performance impact is below 3\% at distances up to 200m. After this point in $\beta$=0.2 and 0.3 veh/m there is an increase in performance of up to 6\%. This is attributable to 3GPP exhibiting a lower mean CBR as shown in Table \ref{table_pdr_densities_aperiodic}, which is most pronounced at higher densities. Aperiodic ETSI traffic performs especially poorly. This is particularly notable given that this represents defined CAM behaviour in accordance with vehicle dynamics as per the standard. It can further be observed that in this case PDR performance does not incur the same sharp decline as others as the density increases. This is because it exhibits similarly low CBR for all densities, as shown in Table \ref{table_pdr_densities_aperiodic}. While more vehicles are transmitting CAMs (125, 250, 400 and 600 vehicles for $\beta$=0.06, 0.12, 0.2 and 0.3 veh/m respectively), the CAM transmission rate decreases as traffic congestion impacts vehicle speed. In our simulations, the mean packet inter-arrival times increases to 122ms, 250ms, 384ms and 610ms for $\beta$=0.06, 0.12, 0.2 and 0.3 veh/m respectively, with ETSI CAMs transmitted less frequently.

\begin{table}[htbp]
\caption{Mean CBR (\%) for varying vehicular densities for periodic and aperiodic application traffic.}
\begin{center}
\resizebox{\columnwidth}{!}{%
\begin{tabular}{|c|c|c|c|c|}
\hline
$\beta$ (veh/m) & Per. (100ms) & Aper. (3GPP) & Aper. (ETSI) & Single Slot\\
\hline
0.06 & 26 & 21 & 19 & 21\\\cline{1-5}
\hline
0.12 & 44 & 37 & 16 & 36\\\cline{1-5}
\hline
0.2 & 67 & 58 & 18 & 49\\\cline{1-5}
\hline
0.3 & 83 & 72 & 15 & 56\\\cline{1-5}
\end{tabular}%
}
\end{center}
\label{table_pdr_densities_aperiodic}
\end{table} 

The degradation in performance for aperiodic traffic can be attributed to frequent grant breaks, leading to inefficient use of SB-SPS resources when application layer packets do not arrive at regular resource reservation intervals. This ultimately results in an increase in packet collisions.

\begin{figure}[htbp]
  \centering
  \includegraphics[width=.82\linewidth]{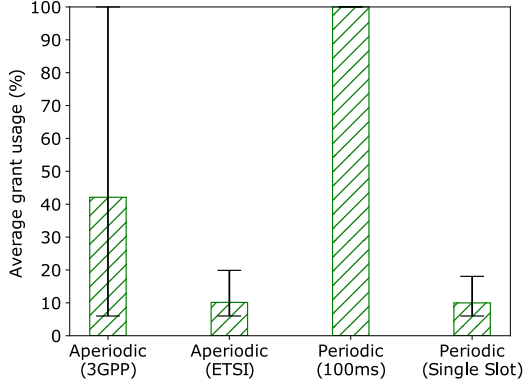}
  \caption{Grant usage for different application models, \(\beta\)=0.12 veh/m.}
  \label{fig:grant-usage}
\end{figure}

The increase in unexpected grant breaks leading to suboptimal grant usage can be observed in Fig. \ref{fig:grant-usage}. When SB-SPS allocates resources in a grant, the re-selection counter dictates the number of reoccurences of said reserved resources i.e. the number of consecutive RRIs. This is randomly selected based on transmission rate, so for periodic CAMs a grant is allocated every 100ms, with the number of reoccurences of the grant randomly allocated in a [5, 15] resource reselection counter (RRC) window. As expected, in Fig. \ref{fig:grant-usage} periodic application traffic maintains 100\% of the allocated grants. By contrast,  aperiodic 3GPP model uses only 42\% of the allocated grants and the aperiodic ETSI model and single slot perform even worse using only 10\% of their respective grants.  This is further investigated in Table \ref{tab:grantInfo} which shows the mean grant length allocated for each application model in accordance with the RRC window, including standard deviation. Table \ref{tab:grantInfo} further shows the mean absolute number of grant transmissions, i.e. used RRIs before the grant is broken or completed, and the mean percentage of grants broken. As previously noted Periodic CAM transmission, results in zero broken grants with all the grant resources being used. The 3GPP aperiodic model breaks 87\% of the grants, while ETSI grant usage mirrors that of single slot usage, where a grant is broken after a single transmission. This represents the worst resource usage, where 100\% of the grants are broken and only 1 reservation is used. 

However, despite only a 13\% difference in the percentage of broken grants between the 3GPP model (87\%) vs the ETSI model (100\%), the 3GPP model demonstrates better performance. This is attributable to better use of the reserved resources before the grant is broken. In Table \ref{tab:grantInfo}, it can be seen that the 3GPP model uses a mean of 4.2 of the allocated resources i.e. RRIs, compared to 1.01 for ETSI. Also the confidence intervals shown in Fig. \ref{fig:grant-usage} and the standard deviation in Table \ref{tab:grantInfo} demonstrate that a considerable amount of the reserved resources are used before the grant is broken. This is because the mean CAM interarrival time for the 3GPP model ranges from 92ms to 100ms across all densities compared to 122ms, 250ms, 384ms and 610ms for the respective densities for the ETSI model (linked to vehicle dynamics). 

\begin{figure}[htbp]
  \centering
  \includegraphics[width=\linewidth]{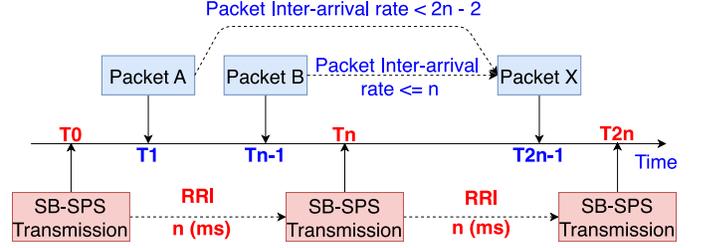}
  \caption{Grant maintenance despite the packet inter-arrival rate exceeding the  resource reservation interval \textit{n}.}
  \label{fig:packetSelection}
\end{figure}

Fig. \ref{fig:packetSelection} shows two scenarios under which a grant can be maintained. In the first scenario, \textit{Packet A} is used to represent the upper time bound between packet transmissions. \textit{Packet A} is generated aperiodically at T\textsubscript{1} for SB-SPS transmission at time T\textsubscript{n}. Thus if the subsequent packet, \textit{Packet X}, is generated at time T\textsubscript{2n-1} this would allow for a packet inter-arrival time of \textit{2n-2} without breaking the grant. In the second scenario, \textit{Packet B} is used to represent the lowest time bound between packet transmissions. \textit{Packet B} is generated aperiodically at T\textsubscript{n-1} for SB-SPS transmission at time T\textsubscript{n}. Thus if \textit{Packet X} is generated at time T\textsubscript{2n-1} this allows a packet inter-arrival time of \textit{n}. As such, if the packet inter-arrival time is \textit{n} or below, the grant will always be maintained. While the packet inter-arrival time is in the range \textit{[n, 2n-2]}, the grant will be maintained if it only traverses one SB-SPS transmission e.g. T\textsubscript{n}. However, if more than one SB-SPS transmission is traversed, the grant will be broken. Thus, given these upper and lower time bounds for the packet inter-arrival rate,  3GPP maintains a rate closer to 100ms and as such maintains the grants for longer. In contrast, ETSI frequently breaks the grant after one transmission due to vehicle dynamics impacting it's packet inter-arrival rates. 

\begin{table}[htbp]
\caption{SB-SPS Grant Usage, \(\beta\)=0.12 veh/m.}
\begin{center}
\resizebox{\columnwidth}{!}{%
\begin{tabular}{|c|c|c|c|}
\hline
{} & Mean Grant Length & Mean Grant Transmissions & Broken Grants \%\\
\hline
Per. (100ms) & $9.87  \pm 3.16$ & $9.87 \pm 3.16$ & 0   \\
Aper. (3GPP)   & $9.982 \pm 3.13$ & $4.20 \pm 3.09$ & 87  \\
Aper. (ETSI)   & $9.98  \pm 3.16$ & $1.01 \pm 0.24$ & 100 \\
Single Slot  & $10.02  \pm 3.16$ & $1.00 \pm 0.00$ & 100 \\
\hline
\end{tabular}%
}
\label{tab:grantInfo}
\end{center}
\end{table}

\begin{figure}[htbp]
\begin{subfigure}{.45\textwidth}
  \centering
  \includegraphics[width=.82\linewidth]{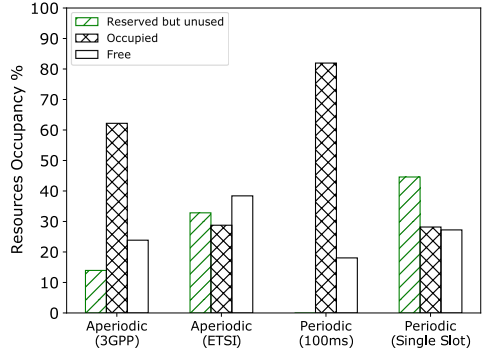}
  \caption{Resource occupancy.}
  \label{fig:distribution-of-resources}
\end{subfigure}
\begin{subfigure}{.48\textwidth}
  \centering
  \includegraphics[width=.82\linewidth]{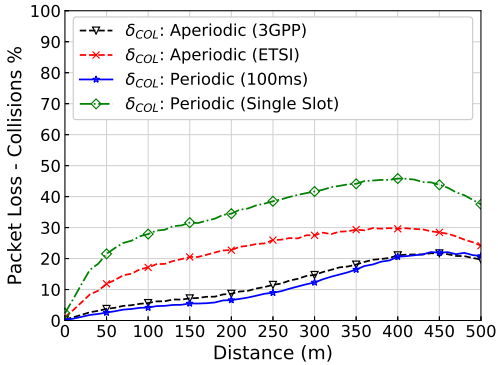}
  \caption{Packet loss due to collisions.}
  \label{fig:medium-scheduling-deltaCol}
\end{subfigure}
\caption{Resource Occupancy \& Packet Collisions for periodic and aperiodic application models, \(\beta\)=0.12 v/m.}
\label{fig:grant-resources}
\end{figure}

The consequence of these grant breaks is an inefficient use of radio resources. A grant break occurs when there is no application layer packet, i.e. TB, to send in a scheduled resource. Thus the resources associated with the scheduled TB, i.e. data packet, go unused. If application layer traffic is highly aperiodic and many grant breaks occur, other vehicles will not be aware that a reserved resource is going unused. Consequently, when choosing possible candidate single subframe resources (CSRs) for their own grant reservations, a VUE will discount the reserved resources from it's deliberations. In dense scenarios, this will lead to each vehicle considering increasingly smaller resource pools, thus increasing the probability that two or more vehicles choose the same free candidate resource, i.e. the same subchannel in the same subframe. This is evident in Fig. \ref{fig:distribution-of-resources} where the resource occupancy is shown for each application model. Resources are classed as follows:

\begin{itemize}
    \item \textbf{Free:} Not used by any VUE.
    \item \textbf{Occupied:} Reserved by a VUE as part of an SB-SPS grant and used to transmit the SCI and TB.
    \item \textbf{Reserved but unused:} Reserved by a VUE as part of an SB-SPS grant but unused as there is no TB available.
\end{itemize}

Fig. \ref{fig:distribution-of-resources} shows that for periodic CAMs and a density \(\beta\)=0.12 veh/m, 82\% of the available radio resources are used with 18\% deemed free. As the 3GPP aperiodic model breaks 87\% of the grants and uses a mean of 42\% of each grant, 14\% of the resources are classed as reserved but unused, i.e. could have been taken but go unused because other vehicles mistakenly believe they're already in use. In the case of the ETSI model the behaviour is different because of the lower CBR, attributable to reduced CAM transmissions based on vehicle dynamics. However, it is still important to observe that the reserved but unused resources exceed those marked as occupied, indicating that resources are mistakenly avoided, even for a low network load. This is further highlighted for the resource occupancy of the worst case model, single slot usage. It should be noted that when calculating resource occupancy, deafness due to distance between vehicles can lead to the same resource being recorded as occupied more than once. This would account for more radio resources than exist in the system and as such we record the resource only once. A resource is recorded as reserved but unused if no TB is sent on it, except in the case where a resource is occupied by 2+ vehicles, where it is recorded just once as occupied.

Ultimately, these frequent grant breaks and inefficient use of resources, coupled with increased vehicle density, leads to a rise in collisions, as VUEs compete for an increasingly small CSR pool, decreasing the overall delivery rate. This is is shown in Fig. \ref{fig:medium-scheduling-deltaCol}.

\section{Radio Resource Allocation for Aperiodic Traffic}
\label{sec:aperiodic_fixes}

We next evaluate how alternative radio resource allocation mechanisms can potentially improve the performance of aperiodic application traffic. This is an open research question and as such the 3GPP does not specify the exact mechanism to be used. The 3GPP Rel. 16 study on NR V2X provides a compilation of proposed approaches \cite{3gpp-TR-38-885}, however, to the best of our knowledge, a thorough analysis and comparison of the performance of said approaches is absent from literature. 
We classify the proposed approaches as those that will work within the confines of the existing SB-SPS mechanism versus those that propose an entirely separate scheduling mechanism for aperiodic application traffic.

\subsection{Aperiodic Scheduling Mechanisms operating within the existing SB-SPS:}
\label{subsec:SB-SPS-approach}
Some of the proposed scheduling mechanisms to support aperiodic traffic still operate within the currently defined SB-SPS mechanism, but modify particular parameters or disregard particular aspects of the scheduling mechanism to improve its performance. The following mechanisms are evaluated:

\begin{itemize}
    \item \textbf{SB-SPS with a reduced sensing window (labeled SW):} The premise of a reduced sensing window is that past reserved sources are not a relevant predictor for the selection of future resources due to the aperiodic packet arrival rate. This resource allocation mechanism was briefly evaluated in \cite{icc-sts}, proposed by Bazzi et. al for CAMs in \cite{Bazzi_2018} and suggested for DENMs in \cite{3gpp-TR-38-885}. In this evaluation we consider shorter sensing windows of \textit{T\textsubscript{s}=500ms}, \textit{200ms} and \textit{100ms}. This is still considered a long term sensing mechanism. 
    \item \textbf{SB-SPS with no RSSI filtering (labeled no RSSI)}\cite{intel-no-rssi}\textbf{:} It is suggested to remove the RSSI filtering stage which typically filters the best 20\% of CSRs ranked by RSSI. This increases the number of CSRs reported to the MAC for potential selection, reducing the potential for collisions due to multiple VUEs selecting the same resource \cite{intel-no-rssi}.  This mechanism can also be classed as a long term sensing.
    \item \textbf{Disabled Grant Breaking:} Grants are generated omitting the \textit{sl-ReselectAfter} parameter, such that the grant will always be maintained despite missed transmissions. This approach is assumed by Molina-Masegosa et. al \cite{aperiodic-molina} but is not compared to a grant breaking scenario.
    \item \textbf{Random Scheduling:} Resources are chosen at random each time a TB is to be transmitted. On packet arrival a single subchannel is selected at random from the subsequent 100 subframes. This is considered the baseline resource allocation mechanism.
\end{itemize}

\begin{figure}[htbp]
\begin{subfigure}{.48\textwidth}
\centerline{
\includegraphics[width=.82\linewidth]{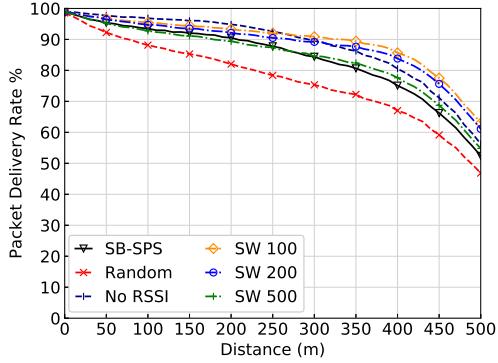}}
\caption{3GPP application model.}
\label{fig:SB-SPS-results}
\end{subfigure}
\begin{subfigure}{.48\textwidth}
\centerline{
\includegraphics[width=.82\linewidth]{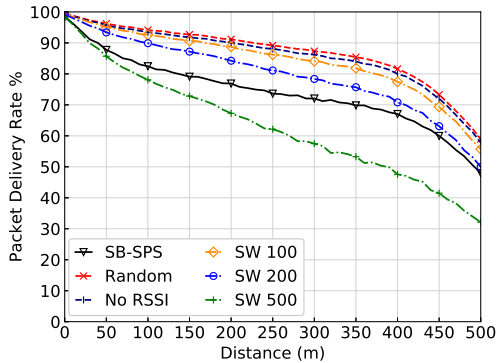}}
\caption{ETSI application model.}
\end{subfigure}
\caption{SB-SPS performance for reduced Sensing Window (SW) duration and removed RSSI filtering for \(\beta\)=0.12 v/m.}
\label{fig:SB-SPS-aperiodic}
\end{figure}

Fig. \ref{fig:SB-SPS-aperiodic} shows the performance of these schemes for the 3GPP and ETSI application models, excluding the disabled grant breaking option which is explored separately. In the case of 3GPP, both removing RSSI filtering and reducing the sensing window results in improved PDR (or comparable performance in the case of a sensing window of 500ms) with all scenarios outperforming the base case of random resource selection. This is because in aperiodic scenarios, historical RSSI filtering which is based on averaging all previous transmissions on the available resources, leads to an inaccurate prediction of how future resources will be utilised. Therefore removing RSSI filtering also increases the randomness of possible CSRs as this stage typically reports only 20\% of possible CSRs, based on lowest RSSI. Removing this filter means that only those packets which are deemed reserved by the RSRP filtering stage will be excluded. In this case, while the proposed CSRs can still be inaccurate as a grant break may have occurred, it is unlikely to cause large amounts of resources to be removed from the CSR pool in lower density scenarios, thus increasing choice. An even larger improvement in performance is shown in the case of the ETSI model (up to a maximum of 14\%), attributable to much lower CBR and clearly showing that the current operation of the SB-SPS mechanism cannot properly deal with aperiodic traffic patterns. 

Reducing the sensing window (SW) duration also shows some improvement for both application models, with a maximum gain of 11\% and 9\% for the 3GPP model, and 12\% and 8\% for the ETSI model for the sensing windows of 100ms and 200ms respectively. In this case the performance is improved by minimising the impact of RSSI filtering on the selection of possible CSRs, with only the most relevant information maintained for the RSRP and RSSI filtering stages. Notably, a sensing window duration of 500ms performs significantly worse than the default 1000ms for the ETSI model and is comparable for the 3GPP model. This highlights a deficiency in using a reduced sensing window as a means of accommodating highly aperiodic traffic, as it needs to be heavily parameterised depending on the characteristics of the application model and density of recorded RSSI values. As 500ms is higher than the lower duration sensing windows, it performs additional, filtering based on RSSI, thereby reducing the number of potential CSRs reported to the MAC layer. By continuation the default 1000ms sensing window should perform worse again. However, this is not the case as there is increased diversity in the CSR selection reported to the MAC by each VUE for 1000ms. For example, for a sensing window of 500ms, the best 20\% of the possible resources may not have recorded values of RSSI and as such are reported directly to the MAC. This results in less diversity in the CSR selection across multiple VUEs and hence increased likelihood of collisions due to choosing the same resource. However for a 1000ms sensing window, only 10\% of resources have no recorded RSSI values with the remaining 10\% chosen based on the sensed RSSI values, leading to greater diversity in the CSR selection across multiple VUEs. However by far the most interesting observation for the ETSI application model, which demonstrates highly aperiodic characteristics, is that while reducing the sensing window and removing RSSI filtering does offer performance improvements, the best performance is achieved by simply randomly allocating the resources. This forms a baseline against which to compare the resource allocation mechanisms evaluated in the next section. 

\begin{figure}[htbp]
\begin{subfigure}{.48\textwidth}
\centerline{
\includegraphics[width=.82\linewidth]{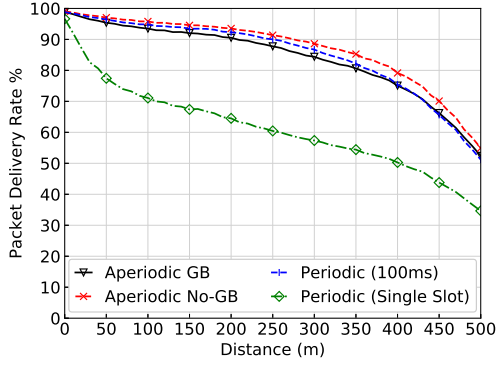}}
\caption{3GPP application model.}
\label{fig:NoGB-SB-SPS-3GPP}
\end{subfigure}
\begin{subfigure}{.48\textwidth}
\centerline{
\includegraphics[width=.82\linewidth]{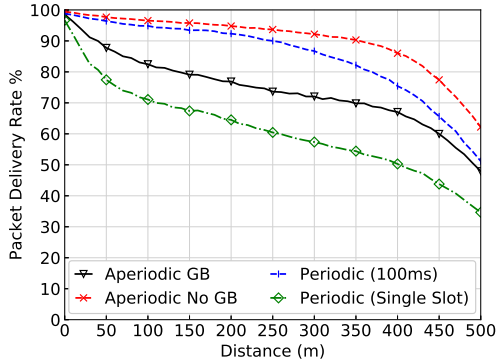}}
\caption{ETSI application model.}
\label{fig:NoGB-SB-SPS-ETSI}
\end{subfigure}
\begin{subfigure}{.45\textwidth}
\centerline{
\includegraphics[width=.82\linewidth]{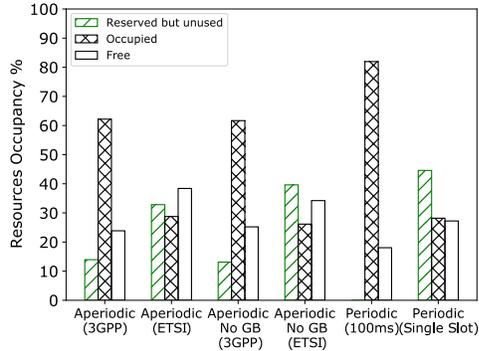}}
\caption{Resource Occupancy.}
\label{fig:NoGB-SB-SPS-ResourceOccupancy}
\end{subfigure}
\caption{SB-SPS performance without grant breaking, \(\beta\)=0.12 v/m.}
\label{fig:NoGB-SB-SPS}
\end{figure}

Fig. \ref{fig:NoGB-SB-SPS} explores the performance of SB-SPS with aperiodic traffic when grant breaking is ignored i.e. the \textit{sl-reselectAfter} parameter is disabled. Maintaining the grant is the default behaviour for the SB-SPS algorithm. Fig. \ref{fig:NoGB-SB-SPS-3GPP} shows the performance of the 3GPP application model with and without grant breaking, where an average PDR increase of 3\% is achieved when grant breaking is disabled. This increase is attributable to a reduction in collisions due to reduced resource rescheduling. In the case of ETSI, as seen in Fig. \ref{fig:NoGB-SB-SPS-ETSI}, the performance increases dramatically, with PDR improvements of up to 20\%. This even out performs periodic traffic due to much lower CBR. Notably, the improvement in performance across both application models derived from disabling grant breaking, comes at a cost. In both cases, the models still have reserved but unused resources. The 3GPP model has 13\% reserved yet unused resources as shown in Fig. \ref{fig:NoGB-SB-SPS-ResourceOccupancy}, which is comparable with a grant breaking scenario. However, the impact is significant in the case of ETSI, where there is an increase of 7\% in the unused resources when introducing disabled grant breaking. There is also an additional source of error, as discussed by Harri et al \cite{cc-hari-missing-sci} when unintentional collisions are introduced. This occurs when a slot goes unused yet the grant is maintained. As an SCI goes unsent neighbouring VUEs may mistakenly believe the resources to be free.

\subsection{Scheduling Mechanisms Designed Specifically for Aperiodic Traffic:}
\label{subsec:nonSB-SPS-approach}
\begin{figure}[htbp]
\begin{subfigure}{.48\textwidth}
    \centerline{
    \includegraphics[width=\linewidth]{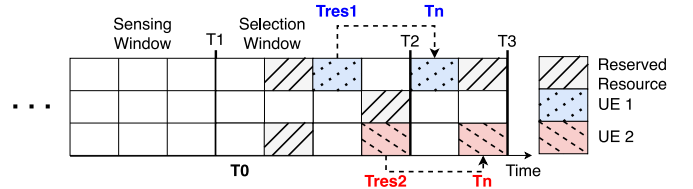}}
    \caption{Short-term reservation Mechanism.}
    \label{fig:one-shot-diagram}
\end{subfigure}
\par\bigskip
\begin{subfigure}{.48\textwidth}
    \centerline{
    \includegraphics[width=\linewidth]{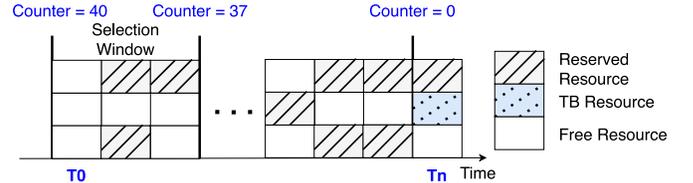}}
    \caption{Counter Based Mechanism.}
    \label{fig:Counter-diagram}
\end{subfigure}
\caption{Scheduling mechanisms for aperiodic traffic.}
\label{fig:Non-SB-SPS-schemes}
\end{figure}

Other scheduling mechanisms have been proposed to specifically support aperiodic traffic and do not work within the existing SB-SPS mechanism. However most could be integrated as a complimentary scheduling mechanism. The mechanisms evaluated are:
\newpage

\begin{itemize}
        \item \textbf{Short-term reservations (labeled STR) \cite{ericsson-one-shot}:} Proposed by Ericsson, short-term reservations also referred to as short term sensing, makes use of two selection windows between T\textsubscript{1}-T\textsubscript{2} and  T\textsubscript{2}-T\textsubscript{3} as shown in Fig. \ref{fig:one-shot-diagram}. It works by sending a reservation signal in the first selection window, at \textit{T\textsubscript{res1}} to reserve resources in the second selection window on which to transmit the SCI and TB pair (\textit{T\textsubscript{n}}). In this mechanism, the resource on which to send the reservation signal is chosen randomly, discounting resources reserved for SCI and TB pairs. Until the reservation signal is sent, the VUE continues to listen in case the reservation slot becomes reserved by another VUE e.g. \textit{T\textsubscript{res2}}. At the time of sending the reservation signal, a future free resource is selected if it has not already been reserved in T\textsubscript{2}-T\textsubscript{3}. This reduces contention for the transmission of SCI and TB pairs. 
        
        \item \textbf{Counter Based Mechanism \cite{lg-counter}:} This approach makes use of a simple counter to increase the randomness of the resource selection process as shown in Fig. \ref{fig:Counter-diagram}. Upon receiving an application layer packet, a counter is randomly selected between \{1, 40\}. In each subframe, the counter is decremented by the number of free subchannels. Once the counter reaches 0, a free subchannel is chosen at random in the next available subframe where the SCI and TB pair will be transmitted. 
        
    \item \textbf{Random Scheduling:} As per the previous section, resources are chosen at random from the subsequent 100 subframes. 
\end{itemize}

\begin{figure}[htbp]
\begin{subfigure}{.48\textwidth}
  \centering
  \includegraphics[width=.82\linewidth]{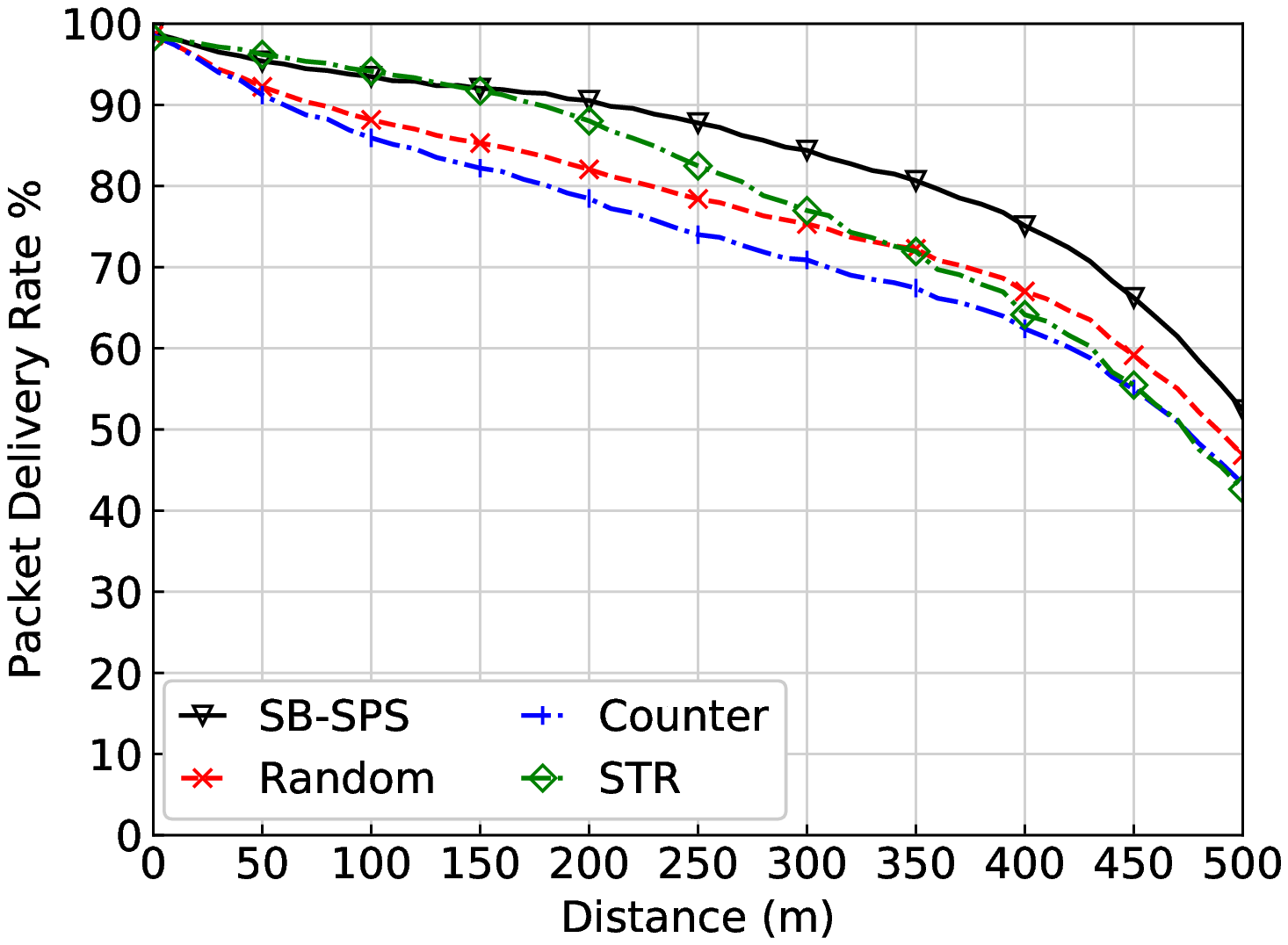}
  \caption{3GPP Application Model.}
  \label{fig:nonSB-SPS-3GPP-results}
\end{subfigure}
\begin{subfigure}{.48\textwidth}
  \centering
  \includegraphics[width=.82\linewidth]{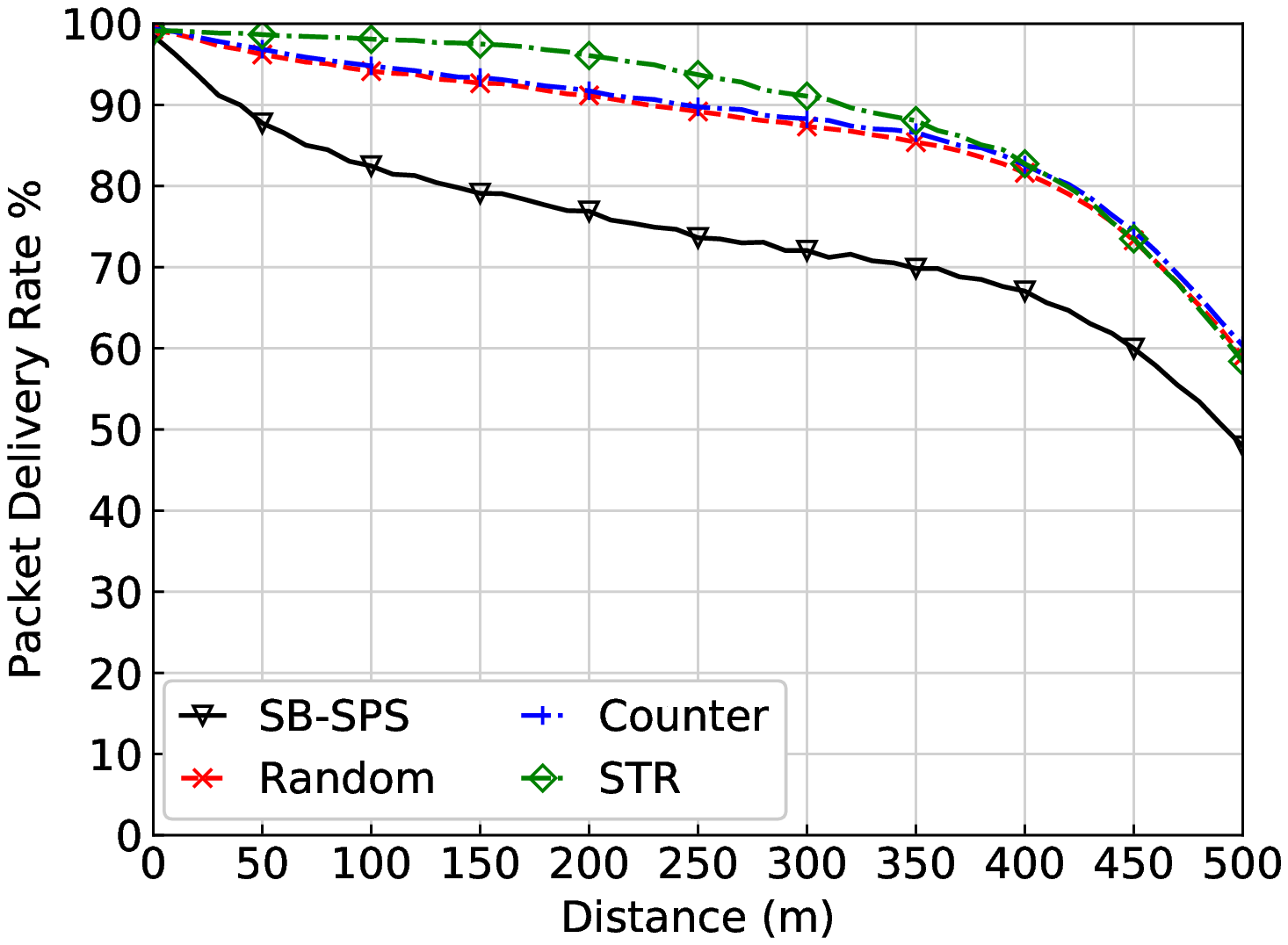}
  \caption{ETSI Application Model.}
  \label{fig:nonSB-SPS-ETSI-results}
\end{subfigure}
\caption{PDR as a function of distance for aperiodic scheduling mechanisms, \(\beta\)=0.12 v/m.}
\label{fig:Non-SB-SPS-results}
  \centering
  \includegraphics[width=.82\linewidth]{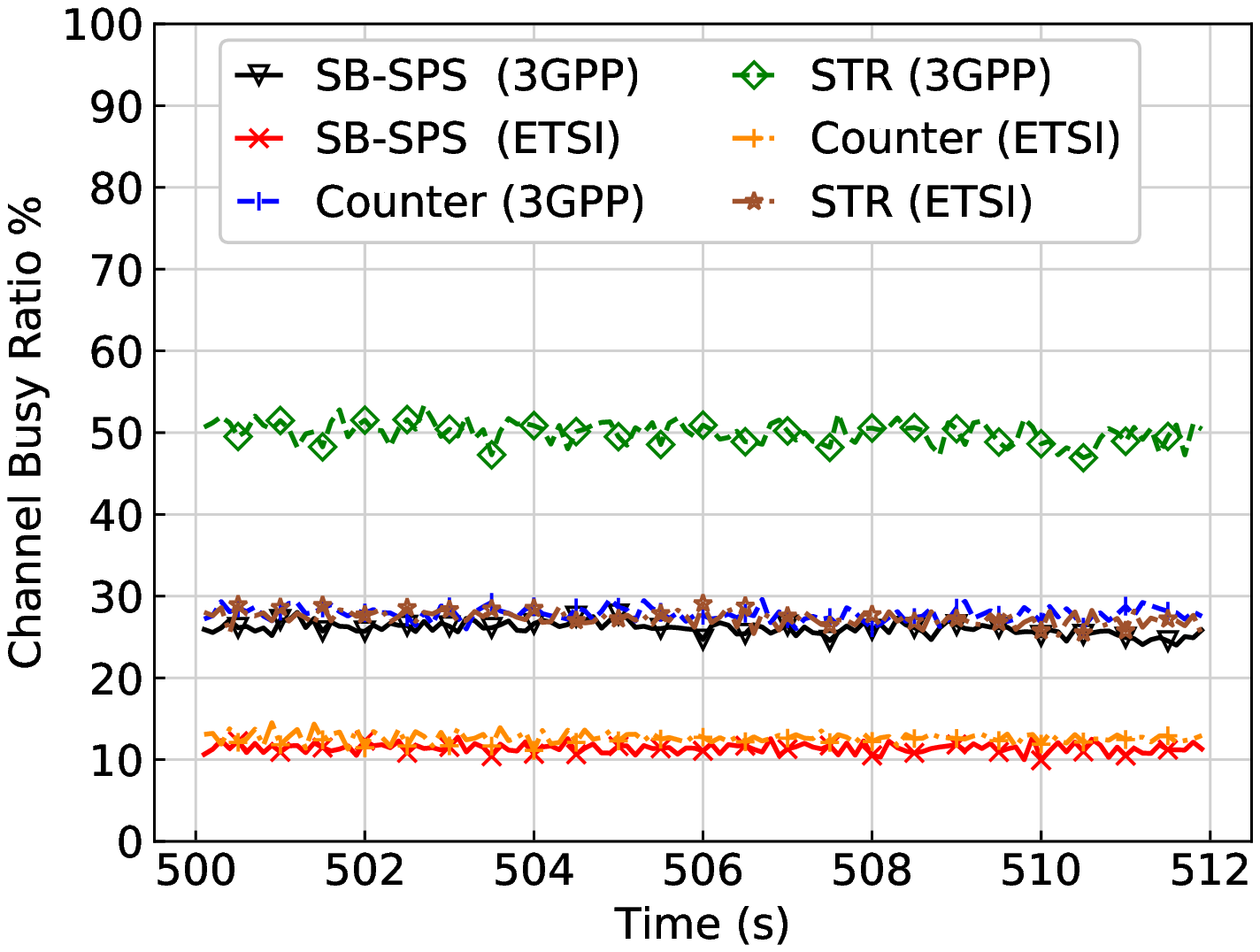}
  \caption{PSCCH Channel Busy Ratio (CBR) Comparison for \(\beta\)=0.12 v/m.}
  \label{fig:oneShot-CBR-PSCCH} 
\end{figure}

Fig. \ref{fig:nonSB-SPS-3GPP-results} shows that the short-term reservation and the counter based mechanisms do not improve the performance of the 3GPP application model, which demonstrates little performance degradation in comparison to the periodic application model due to low divergence in the packet inter-arrival rates (staying in the bounds of \textit{[n, 2n-2]} as already defined). At closer proximities, the STR model performs similarly to SB-SPS but performance quickly drops in accordance with the counter mechanism by 400m. This is due to an increase in congestion of 24\% in the PSCCH as shown in Fig. \ref{fig:oneShot-CBR-PSCCH} caused by the additional reservation signals. 

Fig. \ref{fig:nonSB-SPS-ETSI-results} shows that the short-term reservation and counter based mechanisms considerably improve the PDR performance for ETSI traffic when compared to SB-SPS. This is because these scheduling mechanisms avoid the pitfalls of incorrectly labeling resources as reserved, thereby artificially reducing the available CSRs, with the resource subsequently going unused. It is worth noting that the counter mechanism performs comparably to a simple random allocation approach, so that the additional complexity does not provide much benefit in terms of additional performance. Overall the short-term reservation performs the best, although this is at the expense of additional overhead in the PSCCH due to the  reservation signals, increasing load within the network by approximately 16\% as shown in Fig. \ref{fig:oneShot-CBR-PSCCH}. 

Ultimately, 3GPP application model performance is considerably worse due to the fact that, while aperiodic, it maintains the SB-SPS grant for longer than the ETSI model as per Table \ref{tab:grantInfo} and discussed in Section \ref{sec:aperiodic_traffic}. As discussed the grants are maintained for longer because the 3GPP  packet  inter-arrival  times  often does not  exceed  \textit{2n-2} across a single RRI, where n is the default RRI. Thus, the variability in the packet inter-arrival times strongly impacts which scheduling mechanism is effective. As noted earlier in Fig. \ref{fig:interPacketArrival} the 3GPP inter-arrival times are mostly confined within a small bounded time duration whereas it can be observed that for the ETSI model they are distributed more widely, often exceeding \textit{2n-2}. 

\subsection{Mixed Application Models:}
\label{subsec:mixed_traffic}

Aperiodic scheduling mechanisms such as the counter-based scheduling and short-term reservations can be incorporated to work alongside SB-SPS for mixed traffic scenarios. In both cases, this requires that the aperiodic scheduling mechanisms account for SB-SPS grants when selecting a resource. To evaluate if this approach has any negative impact on the default SB-SPS scheduling, 3 scenarios have been modeled with 10\%, 25\% and 50\% of traffic being aperiodic and conversely 90\%, 75\% and 50\% traffic modeled periodically. 

\begin{figure}[htbp]
\begin{subfigure}{.48\textwidth}
    \centering
    \includegraphics[width=.82\linewidth]{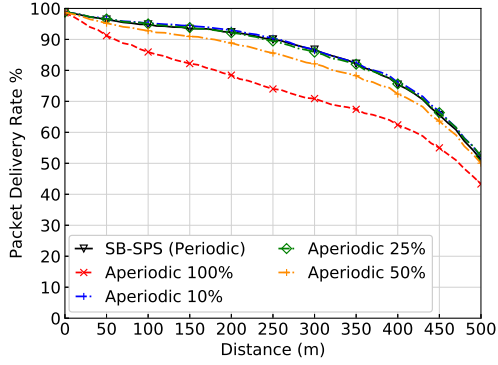}
    \caption{3GPP.}
    \label{fig:mixed-counter-3GPP}
\end{subfigure}
\begin{subfigure}{.48\textwidth}
    \centering
    \includegraphics[width=.82\linewidth]{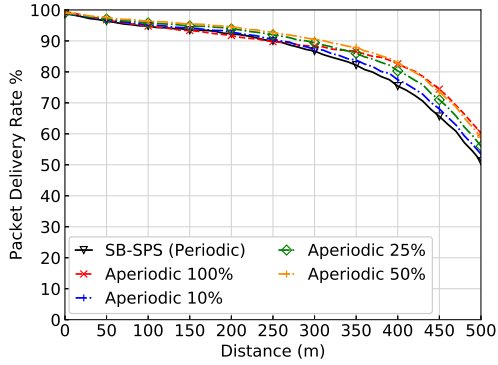}
    \caption{ETSI.}
    \label{fig:mixed-counter-ETSI}
\end{subfigure}
\caption{Mixed application models for the aperiodic Counter Mechanism for \(\beta\)=0.12 v/m.}
\label{fig:Mixed-Counter}
\end{figure}

Figs. \ref{fig:mixed-counter-3GPP} and \ref{fig:mixed-counter-ETSI} show the performance of the counter based mechanism in these mixed traffic scenarios for both 3GPP and ETSI application models. For the ETSI model, it is clear that the mechanism works well in conjunction with the default SB-SPS algorithm, with neither traffic pattern impacting the other in a significant way. In this case the increase in PDR is due to a lowering of CBR when aperiodic traffic is introduced. Similarly, for the 3GPP traffic pattern, the performance is similar to SB-SPS in a purely periodic scenario, which demonstrates that utilising a second aperiodic scheduling scheme such as counter has no impact.

\begin{figure}[htbp]
    \centering
    \includegraphics[width=\linewidth]{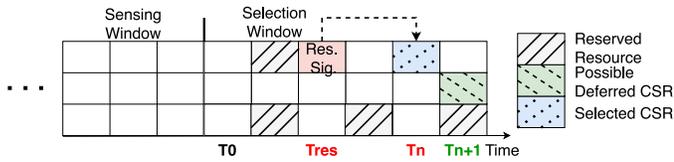}
    \caption{Co-existence of the SB-SPS and short-term Scheduling from the perspective of a VUE.}
    \label{fig:oneShotCoexist}
\end{figure}

Next, we evaluated the performance of the short-term reservation mechanism in mixed traffic scenarios which highlighted a significant issue that would need to be accounted for if these scheduling mechanisms were to co-exist. It is important that the SB-SPS algorithm considers received reservation signals when determining what future resources can be selected as CSRs. This is shown in Fig. \ref{fig:oneShotCoexist}. At \textit{T\textsubscript{0}}, the SB-SPS VUE starts the CSR selection process and selects a future CSR shown in blue (dotted). However to co-exist with the short-term reservation mechanism, it must be ensured that no subsequent reservation signal, received at \textit{TRes}, might reserve the same CSR that has previously been reported to the MAC. To counteract this, SB-SPS must be adapted such that the PHY layer selects the CSR but defers reporting to the MAC layer until the packet needs to be transmitted, thus accounting for reservations such as \textit{TRes} selecting the same resource. In the case that \textit{TRes} chooses the same resource, SB-SPS should relinquish the CSR and make another selection from the top 20\% of CSRs identified in the selection process. 

\begin{figure}[htbp]
    \centering
    \includegraphics[width=.82\linewidth]{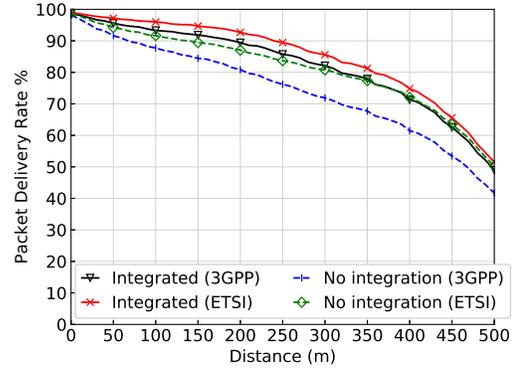}
    \caption{Impact of scheduling co-existence for the 50\% Periodic/Aperiodic mixed application models.}
    \label{fig:oneShotFix}
\end{figure}

The impact of this on the PDR is shown in Fig. \ref{fig:oneShotFix}. A significant decrease in PDR performance is noted of approximately 10\% for 3GPP and ~7\% for the ETSI model. It is worth noting that in our experiments, up to 32000 reservations were made in the 50\% mixed scenario, which led to unnecessary collisions that were avoidable if SB-SPS accounted for the short-term reservations. 

\begin{figure}[htbp]
\begin{subfigure}{.48\textwidth}
  \centering
  \includegraphics[width=.82\linewidth]{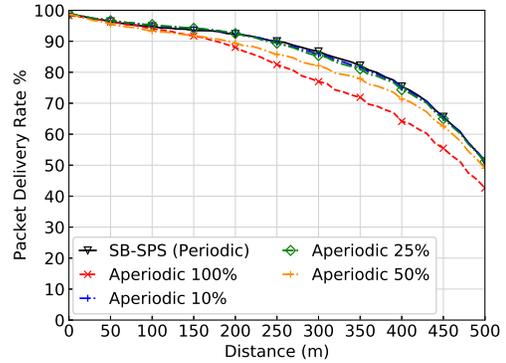}
  \caption{3GPP.}
  \label{fig:Mixed-oneshot-3gpp}
\end{subfigure}
\begin{subfigure}{.48\textwidth}
  \centering
  \includegraphics[width=.82\linewidth]{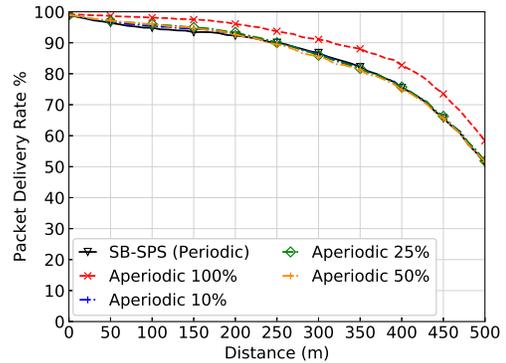}
  \caption{ETSI.}
  \label{fig:mixed-oneShot-etsi}
\end{subfigure}
\caption{Mixed application models for the aperiodic short-term reservation mechanism for \(\beta\)=0.12 v/m (Integration of scheduling applied).}
\label{fig:Mixed-OneShot}
\end{figure}

Assuming this fix is applied to SB-SPS to allow for the scheduling mechanisms to co-exist, Figs. \ref{fig:Mixed-oneshot-3gpp} and \ref{fig:mixed-oneShot-etsi} show the performance of the short-term reservation mechanism with mixed traffic patterns. The pattern follows the results described in earlier sections, showing that the application model has a significant impact on the performance of the mechanism. The increase in load in the PSCCH caused by the short-term reservation signals results in the loss of more SCI messages and consequently TBs. The short-term reservation mechanism can thus be detrimental to both the SB-SPS algorithm and to its own performance in higher density scenarios. When analysing the ETSI model results it can be seen that the short-term reservation mechanism performs comparably with a purely periodic scenario. This demonstrates that for lower channel loads, the short-term reservation mechanism is beneficial to the performance of the aperiodic traffic.

\section{Summary \& Discussion}
\label{sec:discussion}

\begin{figure}[htbp]
\begin{subfigure}{.48\textwidth}
  \centering
  \includegraphics[width=.82\linewidth]{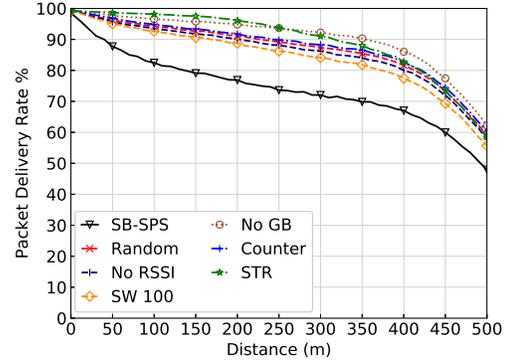}
  \caption{\(\beta\)=0.12 veh/m.}
  \label{fig:ETSI-all}
\end{subfigure}
\begin{subfigure}{.48\textwidth}
  \centering
  \includegraphics[width=.82\linewidth]{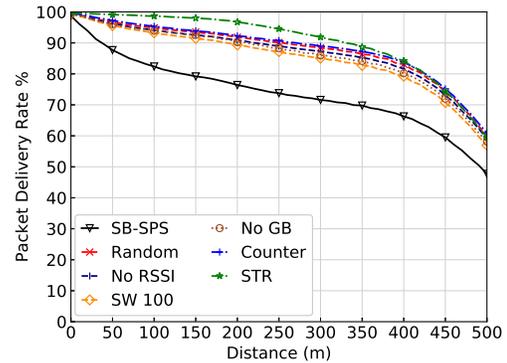}
  \caption{\(\beta\)=0.3 veh/m.}
  \label{fig:etsi-03}
\end{subfigure}
\caption{Summary of PDR Performance for the ETSI Aperiodic Application Model for a variety of scheduling mechanisms.}
\label{fig:etsi-discussion}
\end{figure}

The SB-SPS algorithm was designed with periodic application traffic in mind. Its performance declines sharply when application models exhibit highly aperiodic characteristics, such as those demonstrated by the standardised ETSI CAM traffic generation rules. If packet inter-arrival times are in the range \textit{[n, 2n-2]}, there is a possibility of maintaining the resource grant. However, exceeding this upper bound leads to grant breaks thereby wasting resources and increasing collisions. In Section \ref{subsec:SB-SPS-approach}, we evaluated how carefully parameterising SB-SPS could counter-act this, but as shown in Fig. \ref{fig:ETSI-all} this typically only performs comparably or worse than simply randomly allocating resources. Two schemes that outperform a basic random scheduling approach are STR and disabled grant breaking, achieving improvements of up to 5\% as shown in Fig. \ref{fig:ETSI-all}. However disabled grant breaking demonstrates limited scalability with vehicular density as shown in Fig. \ref{fig:etsi-03}. Despite vehicular densities of \(\beta\)=0.12 and \(\beta\)=0.3 veh/m demonstrating comparable CBR loads due to CAM generation being linked to vehicle dynamics, the increase in packet inter-arrival times results in more reserved yet unused resources. This increases collisions because VUEs mistakenly believe a resource to be free as it has gone unused in previous subframes. In contrast, STR which demonstrates comparable improvement performance benefits to disabled grant breaking at lower vehicular density of \(\beta\)=0.12 veh/m, holds this performance increase at higher vehicular densities. This is as a result of its own dedicated reservation scheme. However, it should be noted that this comes at the cost of added load in the PSCCH due to reservation signals as discussed in Section \ref{subsec:nonSB-SPS-approach}. Ultimately of the schemes evaluated, STR performs best for ETSI CAMs assuming overall channel load is not prohibitive.  

\section{Conclusion}
\label{sec:conclusion}

This paper has presented a full-stack, open source implementation of the 3GPP C-V2X Mode 4 standard, validated to ensure its accuracy. This contribution will enable the vehicular networking community to leverage this work to verifiably investigate the performance of V2X communications from PHY layer up to and including future envisaged application and services. An indepth study on the limitations of SB-SPS to support application traffic with aperiodic characteristics was provided, particularly for the ETSI CAM standard and the 3GPP application model. Mechanisms for improving performance such as parameterisation within the SB-SPS mechanism itself, as well as dedicated aperiodic scheduling mechanisms proposed as potential solutions in future C-V2X standards were evaluated, with the authors highlighting that the level of variability in the packet inter-arrival times have a major impact on the efficacy of these schemes. As such, there is a need to further investigate the communication characteristics of networks with mixed application traffic models to determine how to best schedule resources, while maintaining adequate reliability and quality of service. 

\section*{Acknowledgements}
This publication has emanated from research conducted with the financial support of Science Foundation Ireland (SFI) under Grant No: 17/RC-PhD/3479.

\printbibliography

@ARTICLE{analytical,  author={M. {Gonzalez-Martín} and M. {Sepulcre} and R. {Molina-Masegosa} and J. {Gozalvez}},  journal={IEEE Transactions on Vehicular Technology},   title={Analytical Models of the Performance of C-V2X Mode 4 Vehicular Communications},   year={2019},  volume={68},  number={2},  pages={1155-1166},}

@ARTICLE{3gpp,
author={3GPP},  journal={3GPP, Tech. Rep.},   title={TR 36.885 Study on LTE-based V2X services (v14.0.0, Release 14)},   year={Jul. 2016},}

@ARTICLE{etsi-cam,
 author = {ETSI},
 Title = {EN 302 637-2 Intelligent Transport Systems (ITS); Vehicular Communications; Basic Set of Applications; Part 2: Specification of Cooperative Awareness Basic Service (V1.4.1)},
 Year = {Jan. 2019}
 }

@InProceedings{simulte,
author="Virdis, Antonio
and Stea, Giovanni
and Nardini, Giovanni",
editor="Obaidat, Mohammad S.
and {\"O}ren, Tuncer
and Kacprzyk, Janusz
and Filipe, Joaquim",
title="Simulating LTE/LTE-Advanced Networks with SimuLTE",
booktitle="Simulation and Modeling Methodologies, Technologies and Applications ",
year="2015",
publisher="Springer International Publishing",
address="Cham",
pages="83--105",
isbn="978-3-319-26470-7"
}

@ARTICLE{3gppMeasurements, 
author={3GPP},
journal={3GPP, Tech. Spec.},
title={TR 36.214 Physical layer; Measurements (V14.4.0, Release 14)},
year={Dec. 2017},
}

@ARTICLE{3gppPhyProcedures, 
author={3GPP},
journal={3GPP, Tech. Spec.},
title={TR 36.213 Physical layer procedures (V14.2.0, Release 14)},
year={Dec. 2017},
}

@ARTICLE{Molina-Masegosa,
  author={R. {Molina-Masegosa} and J. {Gozalvez}},
  journal={IEEE Vehicular Technology Magazine}, 
  title={LTE-V for Sidelink 5G V2X Vehicular Communications: A New 5G Technology for Short-Range Vehicle-to-Everything Communications}, 
  year={2017},
  volume={12},
  number={4},
  pages={30-39},}

@inproceedings{artery,
  author    = {Raphael Riebl and
               Hendrik{-}Jorn Gunther and
               Christian Facchi and
               Lars C. Wolf},
  title     = {Artery: Extending Veins for {VANET} applications},
  booktitle = {2015 International Conference on Models and Technologies for Intelligent
               Transportation Systems (MT-ITS), Budapest, Hungary, June 3-5, 2015},
  pages     = {450--456},
  year      = {2015},
  url       = {https://doi.org/10.1109/MTITS.2015.7223293},
  doi       = {10.1109/MTITS.2015.7223293},
  bibsource = {dblp computer science bibliography, https://dblp.org}
}

@INPROCEEDINGS{openCV2XFirst,
  author={B. {McCarthy} and A. {O'Driscoll}},
  booktitle={2019 IEEE 24th International Workshop on Computer Aided Modeling and Design of Communication Links and Networks (CAMAD)}, 
  title={OpenCV2X Mode 4: A Simulation Extension for Cellular Vehicular Communication Networks}, 
  year={2019},
  volume={},
  number={},
  pages={1-6},}

@article{simulte-D2D,
  author    = {Giovanni Nardini and
               Antonio Virdis and
               Giovanni Stea},
  title     = {Simulating device-to-device communications in OMNeT++ with SimuLTE:
               scenarios and configurations},
  journal   = {CoRR},
  volume    = {abs/1609.05173},
  month     = {Sept},
  year      = {2016},
  url       = {http://arxiv.org/abs/1609.05173},
  archivePrefix = {arXiv},
  eprint    = {1609.05173},
  bibsource = {dblp computer science bibliography, https://dblp.org}
}

@ARTICLE{3gpp-TR-36-885,
 author = {3GPP},
 Title = {Study on LTE-based V2X services (v14.0.0, Release 14)},
 Year = {Jul. 2016},
 note = {TR 36.885}
 }

@ARTICLE{3gpp-TR-37-885,
	author = {3GPP},
	title = {Study on evaluation methodology of new Vehicle-to-Everything (V2X) use cases for LTE and NR (v15.3.0, Release 15)},
	Year = {Jul. 2019},
	note = {TR 37.885},
}

@ARTICLE{3gpp-rel15,
	author = {3GPP},
	title = {Release 15 Description;
	Summary of Rel-15 Work Items
 (V15.0.0 , Release 15))},
	year = {Sep. 2019},
	note = {TR 21.915},
}

@ARTICLE{3gpp-rel16,
	author = {3GPP},
	title = {Release 16 Description;
	Summary of Rel-16 Work Items
 (V1.0.0 , Release 16))},
	year = {Dec. 2020},
	note = {TR 21.916},
}

@manual{nist-blers,
 author = {Jian Wang, Richard Rouil },
 title = {BLER Performance Evaluation of LTE Device-to-Device Communications},
 month = {Nov},
 year = {2016},
 note = {Wireless Networks Division Communications Technology Laboratory (NIST)},
}

@manual{lg-counter,
 author = {LG-Electronics},
 title = {Discussion on resource allocation mechanism for NR V2X},
 month = {March},
 year = {2019},
 note= {3GPP TSG RAN WG1 No.96 R1-1901933}
}

@manual{ericsson-one-shot,
 author = {Ericsson},
 title = {Resource allocation procedures for Mode 2},
 month = {March},
 year = {2019},
 note= {3GPP TSG RAN WG1 No.96 R1-1903166}
}

@INPROCEEDINGS{icc-sts,  author={F. {Romeo} and C. {Campolo} and A. {Molinaro} and A. O. {Berthet}},  booktitle={2020 IEEE International Conference on Communications Workshops (ICC Workshops)},   title={Asynchronous Traffic on the Sidelink of 5G V2X},   year={2020},  volume={},  number={},  pages={1-6},}

@article{Bazzi_2018,
   title={Study of the Impact of PHY and MAC Parameters in 3GPP C-V2V Mode 4},
   volume={6},
   ISSN={2169-3536},
   url={http://dx.doi.org/10.1109/ACCESS.2018.2883401},
   DOI={10.1109/access.2018.2883401},
   journal={IEEE Access},
   publisher={Institute of Electrical and Electronics Engineers (IEEE)},
   author={Bazzi, Alessandro and Cecchini, Giammarco and Zanella, Alberto and Masini, Barbara M.},
   year={2018},
   pages={71685–71698}
}

@ARTICLE{3gpp-TR-38-885,
 author = {3GPP},
 Title = {Study on NR Vehicle-to-Everything (V2X) (v16.0.0, Release 16)},
 Year = {Mar. 2019},
 note = {TR 38.885}
 }

@ARTICLE{3gpp-noise,
	author = {3GPP},
	title = {Technical Specification Group Radio Access Network; Evolved Universal Terrestrial Radio Access (E-UTRA); Radio Frequency (RF) requirements for LTE Pico Node B (v16.0.0, Release 16)},
	year = {Jun. 2020},
	year = 2020,
	note = {TR 36.931},
}

@ARTICLE{3gpp-rel12,
	author = {3GPP},
	title = {Technical Specification Group Radio Access Network; Study on LTE Device to Device Proximity Services; Radio Aspects (v12.0.1, Release 12)},
	year = {Mar. 2014},
	year = 2014,
	note = {TR 36.843},
}

@manual{intel-no-rssi-initial,
 author = {Intel},
 title = {Resource Allocation Schemes for NR V2X Sidelink Communication},
 month = {November},
 year = {2018},
 note= {3GPP TSG RAN WG1 Meeting No.95 R1-1812491}
}

@manual{intel-no-rssi,
 author = {Intel Corporation},
 title = {Sidelink Resource Allocation Mechanisms for NR V2X Communication},
 month = {August},
 year = {2018},
 note= {3GPP TSG RAN WG1 Meeting no. 94 R1-1808696}
}

@INPROCEEDINGS{ns3-model,
  author={F. {Eckermann} and M. {Kahlert} and C. {Wietfeld}},
  booktitle={2019 IEEE 90th Vehicular Technology Conference (VTC2019-Fall)}, 
  title={Performance Analysis of C-V2X Mode 4 Communication Introducing an Open-Source C-V2X Simulator}, 
  year={2019},
  volume={},
  number={},
  pages={1-5},
  doi={10.1109/VTCFall.2019.8891534}}

@INPROCEEDINGS{roux-model,  author={P. {Roux} and S. {Sesia} and V. {Mannoni} and E. {Perraud}},  booktitle={2019 IEEE 16th International Conference on Mobile Ad Hoc and Sensor Systems (MASS)},   title={System Level Analysis for ITS-G5 and LTE-V2X Performance Comparison},   year={2019},  volume={},  number={},  pages={1-9},  doi={10.1109/MASS.2019.00010}}

@INPROCEEDINGS{lte-v2vsim,  author={G. {Cecchini} and A. {Bazzi} and B. M. {Masini} and A. {Zanella}},  booktitle={2017 5th IEEE International Conference on Models and Technologies for Intelligent Transportation Systems (MT-ITS)},   title={LTEV2Vsim: An LTE-V2V simulator for the investigation of resource allocation for cooperative awareness},   year={2017},  volume={},  number={},  pages={80-85},  doi={10.1109/MTITS.2017.8005625}}

@ARTICLE{aperiodic-molina,  author={R. {Molina-Masegosa} and J. {Gozalvez} and M. {Sepulcre}},  journal={IEEE Access},   title={Comparison of IEEE 802.11p and LTE-V2X: An Evaluation With Periodic and Aperiodic Messages of Constant and Variable Size},   year={2020},  volume={8},  number={},  pages={121526-121548},  doi={10.1109/ACCESS.2020.3007115}}

@inproceedings{Artery-c,
author = {Hegde, Anupama and Festag, Andreas},
title = {Artery-C: An OMNeT++ Based Discrete Event Simulation Framework for Cellular V2X},
year = {2020},
isbn = {9781450381178},
publisher = {Association for Computing Machinery},
address = {New York, NY, USA},
url = {https://doi.org/10.1145/3416010.3423240},
doi = {10.1145/3416010.3423240},
booktitle = {Proceedings of the 23rd International ACM Conference on Modeling, Analysis and Simulation of Wireless and Mobile Systems},
pages = {47–51},
numpages = {5},
location = {Alicante, Spain},
series = {MSWiM '20}
}

@INPROCEEDINGS{bazzi2020congestion,  author={A. {Bazzi}},  booktitle={2019 53rd Asilomar Conference on Signals, Systems, and Computers},   title={Congestion Control Mechanisms in IEEE 802.11p and Sidelink C-V2X},   year={2019},  volume={},  number={},  pages={1125-1130},  doi={10.1109/IEEECONF44664.2019.9048738}}

@INPROCEEDINGS{icc-repetition,  author={F. {Romeo} and C. {Campolo} and A. {Molinaro} and A. O. {Berthet}},  booktitle={2020 IEEE 91st Vehicular Technology Conference (VTC2020-Spring)},   title={DENM Repetitions to Enhance Reliability of the Autonomous Mode in NR V2X Sidelink},   year={2020},  volume={},  number={},  pages={1-5},  doi={10.1109/VTC2020-Spring48590.2020.9129367}}

@manual{Qualcomm-meeting,
 author = {Qualcomm},
 title = {Sidelink Resource Allocation Mechanism for NR V2X},
 month = {November},
 year = {2018},
 note= {3GPP TSG RAN WG1 Meeting No.95 R1-1813424}
}

@INPROCEEDINGS{cc-hari-missing-sci,
  author={A. {Mansouri} and V. {Martinez} and J. {Härri}},
  booktitle={2019 15th Annual Conference on Wireless On-demand Network Systems and Services (WONS)}, 
  title={A First Investigation of Congestion Control for LTE-V2X Mode 4}, 
  year={2019},
  volume={},
  number={},
  pages={56-63},
  doi={10.23919/WONS.2019.8795500}}

\vspace{-4\baselineskip}

\begin{IEEEbiography}[{\includegraphics[width=1in,height=1.25in,clip,keepaspectratio]{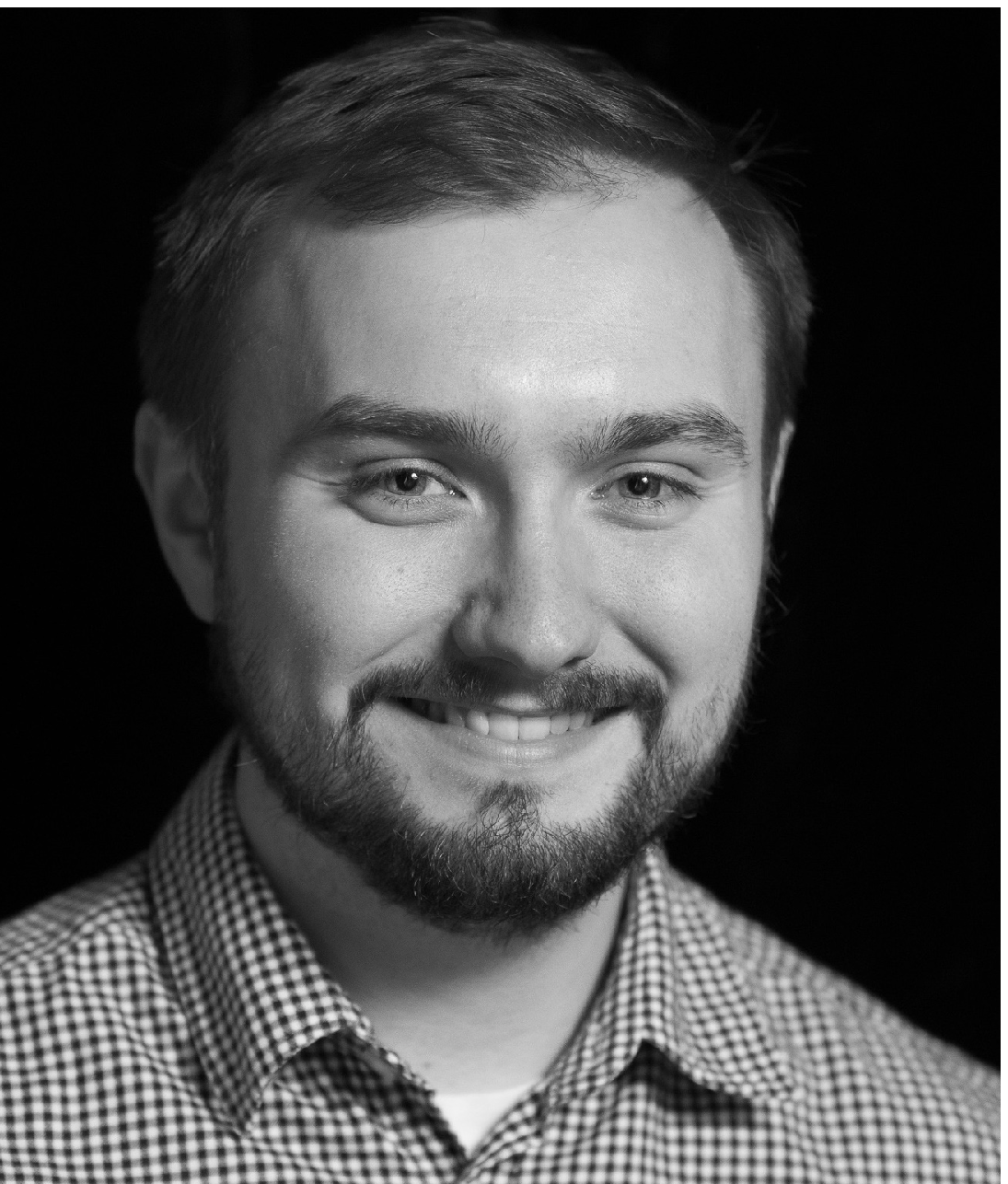}}]%
{Brian McCarthy}
received a B.Sc. degree in Computer Science from University College Cork (UCC), Cork, Ireland, in 2017. He is currently working toward a Ph.D. degree and his research interests include mobile networks, vehicular communications, radio resource management and congestion control.
\end{IEEEbiography}

\vspace{-4\baselineskip}

\begin{IEEEbiography}[
{\includegraphics[width=1in,height=1.25in,clip,keepaspectratio]{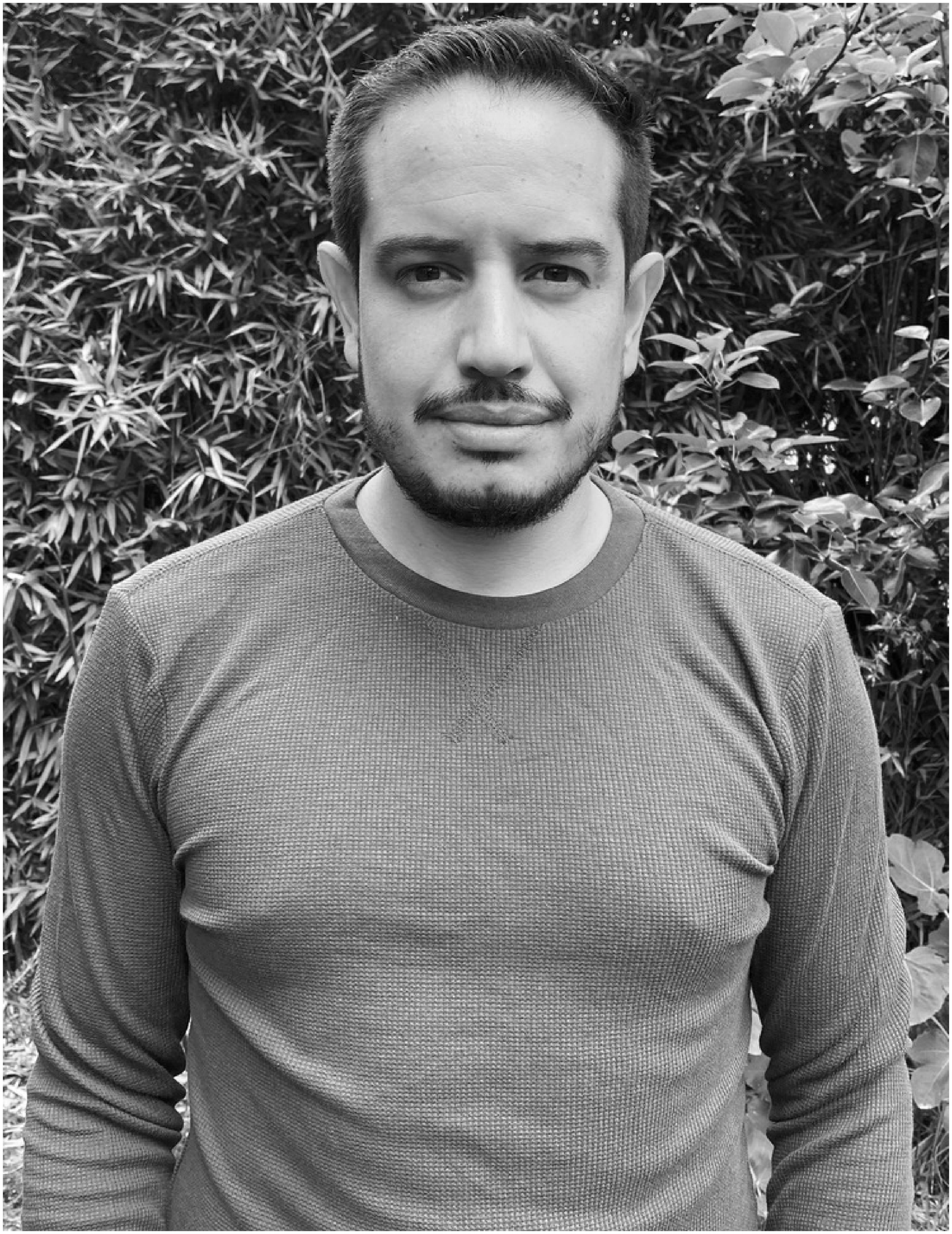}}]%
{Andres Burbano-Abril} received a B.Sc. degree in Electronic Engineering from the University of Azuay (UDA) in Ecuador and the M.Sc. degree in Electrical Engineering (Telecommunications) from the National Autonomous University of Mexico (UNAM) in 2013 and 2016, respectively. His current research focuses on modeling and optimization of 5G-enabled vehicular networks. He is currently a Ph.D. candidate at UNAM.
\end{IEEEbiography}

\vspace{-4\baselineskip}

\begin{IEEEbiography}[
{\includegraphics[width=1in,height=1.25in,clip,keepaspectratio]{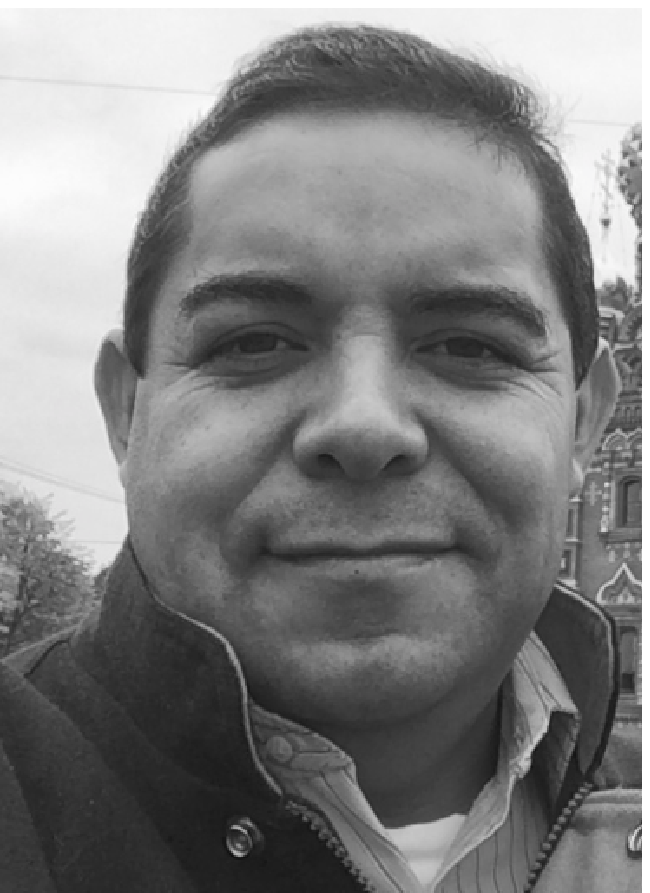}}]%
{Victor Rangel-Licea} received the B.Sc. (Hons.) degree in computer engineering from the National Autonomous University of Mexico (UNAM) in 1996. He received the MSc. in Telematics and Ph.D. in Electrical Engineering degrees (at the C4MCR) from the University of Sheffield, United Kingdom, in 1998 and 2002, respectively. His research focuses on IoT performance analysis, QoS, traffic modeling, and scheduling algorithms for 4G and 5G cellular and vehicular networks. He has been with the School of Engineering at UNAM since 2002 as a research professor.
\end{IEEEbiography}

\vspace{-4\baselineskip}

\begin{IEEEbiography}[
{\includegraphics[width=1in,height=1.25in,clip,keepaspectratio]{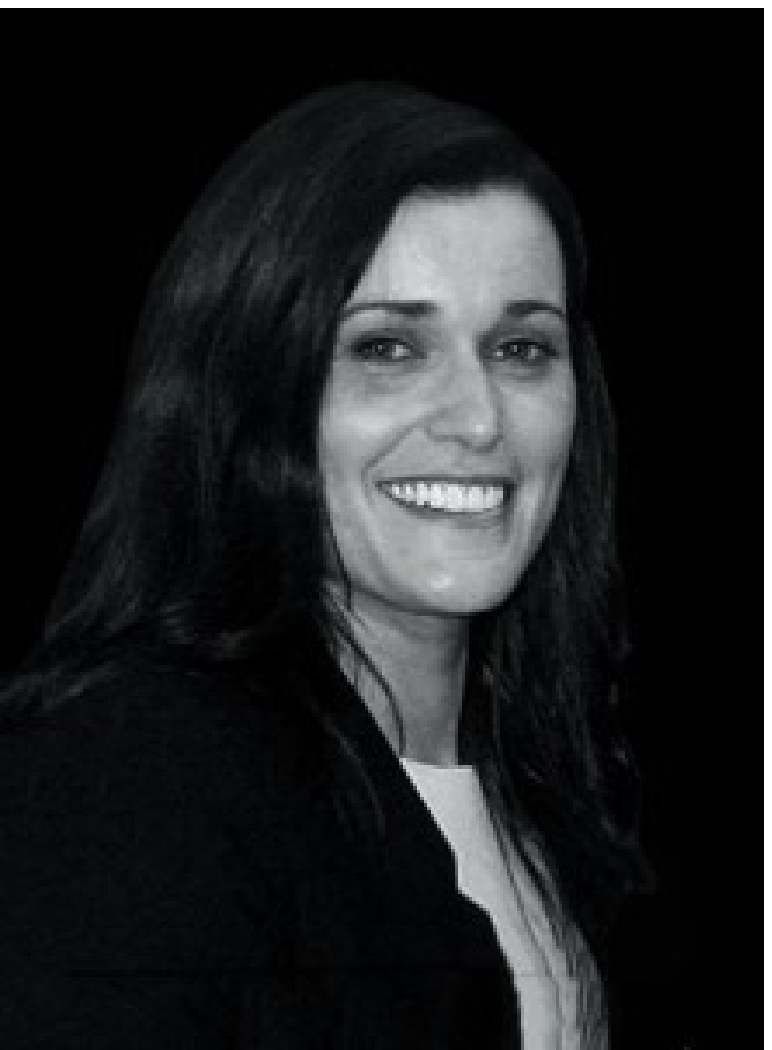}}]%
{Aisling O’Driscoll} received a B.Sc. degree in Software Development and Computer Networking and a M.Eng by research from the Cork Institute of Technology (now Munster Technological University), Cork, Ireland in 2004 and 2006, respectively. She received a Ph.D. from Cork Institute of Technology, in conjunction with the Nimbus Centre for Research in Embedded Systems in 2014. She is a lecturer in the School of Computer Science and Information Technology at University College of Cork (UCC), Ireland, and is a funded investigator in the Science Foundation Ireland (SFI) CONNECT Centre for future networks and telecommunications and Lero, the Irish software research centre. Her research interests include wireless and mobile networks, communication protocols and network management with a particular interest in the vehicular networking domain.
\end{IEEEbiography}

\end{document}